\documentclass[iop,apj,tighten]{emulateapj}
\usepackage{apjfonts,graphicx}
\usepackage{amsmath,amstext}
\usepackage[breaklinks,colorlinks,citecolor=blue,linkcolor=magenta]{hyperref}
\usepackage[all]{hypcap}
\DeclareMathOperator{\erf}{erf}

\newcommand{\Msun}{\,M_{\odot}}
\newcommand{\epsff}{\epsilon_{\mathrm{ff}}}
\newcommand{\epsint}{\epsilon_{\mathrm{int}}}
\newcommand{\taumax}{\tau_{\mathrm{max}}}
\newcommand{\tauave}{\tau_{\mathrm{ave}}}
\newcommand{\tauspread}{\tau_{\mathrm{spread}}}
\newcommand{\fboost}{f_{\mathrm{boost}}}
\newcommand{\Rgmc}{R_{\mathrm{GMC}}}
\newcommand{\SigmaSFR}{\Sigma_{\mathrm{SFR}}}

\begin{document}

\slugcomment{Submitted to ApJ}
\shortauthors{Li et al.}
\shorttitle{Star Cluster Formation in Cosmological Simulations}

\title{Star cluster formation in cosmological simulations.  II. Effects of Star Formation Efficiency and Stellar Feedback}
\author{Hui Li\altaffilmark{1,2*}\href{https://orcid.org/0000-0002-1253-2763}{\includegraphics[scale=0.12]{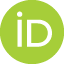}}, Oleg Y. Gnedin\altaffilmark{1}\href{https://orcid.org/0000-0001-9852-9954}{\includegraphics[scale=0.12]{orcid.png}}, Nickolay Y. Gnedin\altaffilmark{3,4,5}}

\altaffiltext{1}{Department of Astronomy, University of Michigan, Ann Arbor, MI 48109, USA}
\altaffiltext{2}{Department of Physics, Kavli Institute for Astrophysics and Space Research, Massachusetts Institute of Technology, Cambridge, MA 02139, USA}
\altaffiltext{3}{Particle Astrophysics Center, Fermi National Accelerator Laboratory, Batavia, IL 60510, USA}
\altaffiltext{4}{Kavli Institute for Cosmological Physics, University of Chicago, Chicago, IL 60637, USA}
\altaffiltext{5}{Department of Astronomy \& Astrophysics, University of Chicago, Chicago, IL 60637, USA}
\altaffiltext{*}{hliastro@mit.edu}

\date{\today}

\begin{abstract}
The implementation of star formation and stellar feedback in cosmological simulations plays a critical role in shaping galaxy properties. In the first paper of the series, we presented a new method to model star formation as a collection of star clusters. In this paper, we improve the algorithm by eliminating accretion gaps, boosting momentum feedback, and introducing a subgrid initial bound fraction, $f_i$, that distinguishes cluster mass from stellar particle mass. We perform a suite of simulations with different star formation efficiency per freefall time $\epsff$ and supernova momentum feedback intensity $\fboost$. We find that the star formation history of a Milky Way-sized galaxy is sensitive to $\fboost$, which allows us to constrain its value, $\fboost\approx5$, in the current simulation setup. Changing $\epsff$ from a few percent to 200\% has little effect on global galaxy properties. However, on smaller scales, the properties of star clusters are very sensitive to $\epsff$. We find that $f_i$ increases with $\epsff$ and cluster mass. Through the dependence on $f_i$, the shape of the cluster initial mass function varies strongly with $\epsff$. The fraction of clustered star formation and maximum cluster mass increase with the star formation rate surface density, with the normalization of both relations dependent on $\epsff$. The cluster formation timescale systematically decreases with increasing $\epsff$. Local variations in the gas accretion history lead to a 0.25~dex scatter for the integral cluster formation efficiency. Joint constraints from all the observables prefer the runs that produce a median integral efficiency of 16\%.
\end{abstract}

\keywords{galaxies: formation --- galaxies: star clusters ---  globular clusters: general}

\section{Introduction}

With the advance of observational discoveries, such as the anisotropy of the cosmic microwave background \citep[e.g.][]{komatsu_etal11, planck15} and the large-scale galaxy clustering \citep[e.g.][]{geller_etal89, bond_etal96, gott_etal05}, the $\Lambda$CDM cosmology and hierarchical structure formation framework have been widely accepted and served as the starting point of theoretical investigations of galaxy formation. Among all theoretical methodologies, numerical simulations have become the main tool to study the formation and evolution of galaxies \citep[see][]{somerville_dave15}.

Built upon the cosmological $N$-body simulations that explored the growth of large-scale structure under only gravity \citep[e.g.][]{davis_etal85, navarro_etal97, governato_etal99, springel_etal05, boylan_kolchin_etal09, stadel_etal09, klypin_etal11}, recent cosmological hydrodynamical simulations with state-of-the-art suites of physical ingredients and numerical techniques started to reproduce various stellar and gaseous properties of the observed galaxies, such as the stellar mass-halo mass relation, the average star formation histories (SFHs), and Kennicutt-Schmidt relations (KSRs), not only at $z\approx 0$ but also at high redshifts \citep[e.g.][]{agertz_etal13, hopkins_etal14, vogelsberger_etal14, schaye_etal15}. This success is achieved partly with the accurate numerical treatments of complex baryonic physics, such as gravity, gas dynamics, and radiative heating and cooling, but is largely due to novel implementations of subgrid models that describe the star formation and stellar feedback processes that cannot be spatially or temporally resolved in these simulations \citep[e.g.][]{cen_ostriker_92, katz92, navarro_white93, katz_etal96, springel_hernquist03}.

During the last two decades, great efforts have been made to explore the sources and implementations of stellar feedback processes in cosmological simulations \citep{stinson_etal06, governato_etal07, scannapieco_etal08, agertz_etal11, agertz_etal13, guedes_etal11, aumer_etal13, booth_etal13, ceverino_etal14, keller_etal14, roskar_etal14}. These works have demonstrated the important role played by the energetic feedback to suppress star formation at high redshift and launch galactic winds \citep[e.g.][]{muratov_etal15}. However, it is troublesome that a surprisingly broad range of feedback models claims to match the same global properties of galaxies by fine-tuning parameters of feedback prescriptions, reducing the predictive power of galaxy formation modeling \citep{naab_ostriker17}. Yet it is still unknown whether the implementations used in these simulations are appropriate for capturing the physical properties of the interstellar medium (ISM) on smaller scales. As the spatial resolution of current simulations is approaching the scales of individual star-forming regions \citep[e.g.][]{hopkins_etal14, read_etal16, wetzel_etal16}, it is critical to develop systematic methods to calibrate the subgrid models on a similar scale.

In \citet[][hereafter Paper~I]{li_etal17} we introduced a new prescription for modeling star formation by considering star clusters as a unit of star formation, following the general consensus that most stars form in cluster environments \citep{lada_lada03}. In this prescription, a cluster particle grows continuously through gas accretion from its natal giant molecular cloud (GMC). The growth of a cluster particle is resolved in time and is terminated by its own energy and momentum feedback. Thus, the final particle mass is set self-consistently and can be considered as the mass of a single star cluster formed within the GMC. Since the cluster growth is determined by both the efficiency of star formation and the strength of stellar feedback, comparing key properties of model clusters with observations provides us with a unique opportunity to constrain these subgrid models on scales of cluster-forming regions, instead of kpc scales.

Recent observations of star-forming regions in the Milky Way and other nearby galaxies reveal a large number of cluster samples that contain the information about their formation environment \citep{portegies_zwart_etal10}. Star clusters follow a well-defined initial mass function (CIMF) that can be described by the Schechter function with a power-law slope of $\approx -2$ and an exponential cutoff at high-mass end. Indeed, as we have shown in Paper~I, the slope of the CIMF reflects the slope of the density distribution of the star-forming gas, modulated by the feedback effects. Moreover, the high mass cutoff is also related to the intensity of star formation activity of the host galaxies \citep[e.g.][]{larsen_02, adamo_etal15, johnson_etal16}.

For a long time, it has been believed that the galaxy-wide low star formation efficiency was caused by the delay of gravitational collapse by magnetic or turbulent support \citep{krumholz_mckee05, krumholz_tan07}. However, recent observations reveal a very short age spread of stars, a few Myr or a couple of freefall timescales, in many young star clusters \citep{maclow_klessen_04, hartmann_etal12, hollyhead_etal15}, suggesting that cluster formation is a rapid and dynamical process. The cluster formation timescale is determined by both the speed of gas accretion and the intensity of stellar feedback, thus providing an additional test of the star formation and feedback implementation in the simulations.

In this paper, we further revise and improve the cluster formation model. We describe the major updates to the model in Section~\ref{sec:updates}, and present results from a new set of simulations in Section~\ref{sec:results}. We perform simulations with a wide range of subgrid model parameters: local star formation efficiency and a boost of supernova (SN) momentum feedback. The CIMF, the fraction of star formation in bound clusters, and the age spreads present new constraints on the subgrid parameters, which we discuss in Section~\ref{sec:discussion}. We summarize our results and conclusions in Section~\ref{sec:summary}.

\section{Simulations}\label{sec:updates}
\subsection{Overview of cluster formation}

The full description of the simulation setup is presented in Paper~I. Here we first briefly summarize the common aspects and then describe the improvements to the model. Most of the simulations in this paper are new and have distinct features from those discussed in Paper~I.

The simulations are run with the Eulerian gas dynamics and $N$-body Adaptive Refinement Tree (ART) code \citep{kravtsov_etal97,kravtsov99,kravtsov03,rudd_etal08}, which includes several key physical ingredients, such as three-dimensional radiative transfer \citep{gnedin_abel01,gnedin14}, non-equilibrium chemical network of hydrogen and helium, non-equilibrium cooling and photoionization heating, phenomenological molecular hydrogen formation and destruction \citep{gnedin_kravtsov11}, and a subgrid-scale turbulence model \citep{semenov_etal16}. We run the cosmological simulations from the initial condition that contains a main halo with the total mass $M_{200} \approx 10^{12}\Msun$ at $z=0$, in a periodic box of 4 comoving Mpc in size. All simulations start on a $128^3$ root grid, which sets the dark matter particle mass $m_{\rm DM} = 1.05 \times 10^{6}\Msun$. The high dynamic range of spatial resolution is achieved by adaptive mesh refinement, where additional cell levels are added to the root grid. We apply a Lagrangian refinement criterion for both dark matter and gas components so that the mass of all gas cells varies only within a narrow range at all times. In addition, we adopt a Jeans refinement criterion with which cells larger than twice the local Jeans length will be refined.

In Paper~I, we have developed a novel algorithm for the formation of star clusters in cosmological simulations: continuous cluster formation (CCF). This algorithm is different from most of the star formation methods implemented in current mesh- and particle-based cosmological simulations, where star particles are spawned instantaneously by a Poisson process with their masses set beforehand. In CCF, each star particle represents a single star cluster formed at a local density peak of the molecular gas. Cluster particles grow their masses continuously by accreting ambient material at every local timestep ($\sim 1000$~years), meaning that the star formation process is resolved with high time resolution. The growth of a cluster particle is terminated by its own energy and momentum feedback. Thus, the final particle mass is set self-consistently and represents the mass of a single star cluster formed within the natal GMC.

In order to avoid dependence on the time-variable physical size of a cell, $L_{\rm cell}$, the cluster particle is allowed to grow its mass via gas accretion within a spherical region of fixed physical size. We interpret this sphere as the dense part of a GMC that would form a bound stellar system in freefall collapse. The optimal value of the sphere size was investigated in Paper I, $\Rgmc=5\,$pc, which is similar to the sizes of observed massive cluster-forming clouds or clumps \citep[e.g.][]{urquhart_etal14}. We will henceforth refer to this star-forming sphere as the "GMC".

At the highest refinement level, the GMC sphere fully includes the peak-density ("central") cell and partially overlaps with its 26 neighbor cells from a $3\times 3\times 3$ cube configuration. Accessing more than these immediate neighbors is computationally prohibitive in the ART code. The growth rate of a given cluster depends on the H$_2$ density of the overlapping cells,
\begin{equation}
  \dot{M} = \sum_{\mathrm{cell}}{f_{\rm GMC}\, V_{\mathrm{cell}}\, \dot{\rho}_{*}} 
  = \frac{\epsff}{\tau_{\rm ff}}\, \sum_{\mathrm{cell}}{f_{\rm GMC}\, V_{\mathrm{cell}}\, f_{\rm H_2}\, \rho_{\rm gas}},
  \label{eq:sfr}
\end{equation}
where $V_{\mathrm{cell}}$ is the volume of each neighbor cell, $f_{\rm GMC}$ is the fraction of $V_{\mathrm{cell}}$ that overlaps with the GMC, $f_{\rm H_2}$ is the mass fraction of hydrogen in molecular phase, $\rho_{\rm gas}$ is the total gas density, and $\epsff$ is the star formation efficiency per freefall time. The freefall time is defined as
\begin{equation}
  \tau_{\rm ff} \equiv \left(\frac{32G\rho}{3\pi}\right)^{-1/2}
  \approx 1.6\,{\rm Myr}\, \left(\frac{n_{\rm H}}{10^3 \,{\rm cm}^{-3}} \right)^{-1/2},
\end{equation}
where $n_{\rm H}$ is the total number density of hydrogen of the central cell. The mass growth of the cluster particle is calculated and added to the particle mass at each local timestep, typically $\Delta t\sim 100$ yr. The mass accumulation history of each cluster is thus temporally resolved with many thousands of steps.

Cluster formation is allowed only in cells above the number density $n_{\rm H,th} = 10^3\,$cm$^{-3}$. This is close to the observational estimate by \citet{kainulainen_etal14} of the H$_2$ density threshold for star formation in nearby GMCs of $\sim 5\times 10^3\,$cm$^{-3}$. In addition, to avoid the creation of very small and numerous clusters, new particles are created only if their expected final mass is above threshold $M_{\rm th}$. The expected mass is calculated as the initial rate $\dot{M}$ times the maximum allowed formation time, $\taumax = 15\,$Myr. Since the actual duration of cluster formation is significantly shorter in the new runs, we discuss and revise $M_{\rm th}$ in Section~\ref{sec:noremoval}.

\subsection{Improvements to Paper~I methodology}

\subsubsection{Gas cell refinement} \label{sec:refinement}

In Paper~I, we employed quasi-Lagrangian refinement criteria, which keep the mass of all cells (except for cells at the finest level) within a similar range. We examined the influence of $\Rgmc$ on CIMF by varying $\Rgmc$ from 2.5 to 7.5 pc and found the best match to observations when the physical size of the gas cells ($L_{\rm cell}$) involved in star formation is comparable to the size of the GMC sphere. Therefore, we tune the refinement strategy such that the physical size of the gas cells at the finest refinement level remains around $R_{\rm GMC}=5\,$pc. Initially, we allow nine refinement levels on the $128^3$ root grid at high redshift until $z\approx9$. As simulations run toward lower redshift, the $10^{\rm th}$, $11^{\rm th}$, and $12^{\rm th}$ refinement levels are added at $z\approx$ 9, 4, and 1.5, respectively. This refinement method keeps the $L_{\rm cell}$ of the finest level in the range between 3 and 6~pc at all cosmic times of interest in this paper.

\subsubsection{Redshift-independent star formation efficiency} \label{sec:eps_ff}

The mass accretion rate of a cluster is given by Equation~(\ref{eq:sfr}), which contains the parameter $\epsff$. However, the meaning of $\epsff$ is quantitatively different from what is commonly used in galaxy formation simulations. In the traditional prescription, stellar particles are given the mass calculated as the star formation rate density $\dot{\rho}_{*}$ times the volume of the cell containing the particle. The cell size is fixed in comoving coordinates but expands in proper physical coordinates, so that for the same gas density (typically, near the threshold for star formation), the particle mass increases with time. In contrast, in our model, each star cluster grows its mass over several million yr by accreting material within a cloud of fixed physical radius, $\Rgmc$. Our $\epsff$ is applied to a fixed volume and does not vary with cosmic time. This is an important improvement.

We can relate our value of $\epsff$ to that on the scale of a cell via the differences in volume. Consider the case when the diameter of the GMC sphere is smaller than a star-forming cell, and therefore the GMC is completely embedded in the cell. The smaller volume allowed to participate in star formation translates into the smaller efficiency for the whole cell,
\begin{equation}
  \epsilon_{\rm ff,cell} = \frac{V_{\rm GMC}}{V_{\rm cell}} \, \epsff
  = \frac{4\pi}{3} \left(\frac{\Rgmc}{L_{\rm cell}}\right)^3 \, \epsff.
\end{equation}
For example, the seemingly high value $\epsff=50\%$ in our CCF model is equivalent to $\epsilon_{\rm ff,cell}\approx 1\%$ for a cell of 30~pc, a typical size for current highest-resolution cosmological simulations.

We allow cluster formation at the three finest refinement levels, which is consistent with our adopted star formation density threshold: $n_{\rm crit} = 10^3\,\mathrm{cm}^{-3}$. For perfectly Lagrangian refinement, the average number density of hydrogen plus helium atoms for cells at level $l$ and redshift $z$ in our simulation box is
\begin{equation}
  n(z) \approx 900\, \mathrm{cm}^{-3} \times 8^{l-9}\left(\frac{1+z}{3}\right)^3.
\end{equation}
We can see that at $z=2$, cells at level 9 (the third-finest level at that time) have an average density close to $n_{\rm crit}$. Cells create new clusters as soon as they exceed the density threshold, which usually happens already at the third-finest level. The physical cell size at that level is in the range $L_{\rm cell}=12-24$~pc, which is larger than $2\Rgmc$. Thus, the corresponding cell efficiency initially is $\epsilon_{\rm ff,cell} = (0.038-0.30)\, \epsff$.

This is an important point for the comparison with the low efficiency per freefall time inferred in Galactic star forming regions, $\epsff \sim 1\%$ \citep{kennicutt98, krumholz_etal12, vutisalchavakul_etal16}. In our model, clusters begin forming with a similarly low efficiency. As the gas continues to collapse in free fall, the density increases and cells are refined to the highest level. Then the effective $\epsilon_{\rm ff,cell}$ increases as well, to match $\epsff$, in agreement with the results of \citet{murray11} and \citet{lee_etal16}. 

As in every numerical simulation, at the highest refinement level, we are not properly resolving gas collapse and therefore underestimate the density and overestimate the freefall time. This numerical effect also requires adopting a larger value of $\epsff$ to maintain the correct star formation rate (SFR).

These algorithmic and numerical issues make it difficult to directly compare our parameter $\epsff$ with the observed efficiency on $\sim 30$~pc scales of GMCs. We will discuss instead the distribution of the integral efficiency, which is the fraction of gas mass turned into stars, in Figure~\ref{fig:epsint} and Section~\ref{sec:all}. 

\begin{figure}[t]
\centerline{
\includegraphics[width=0.9\hsize]{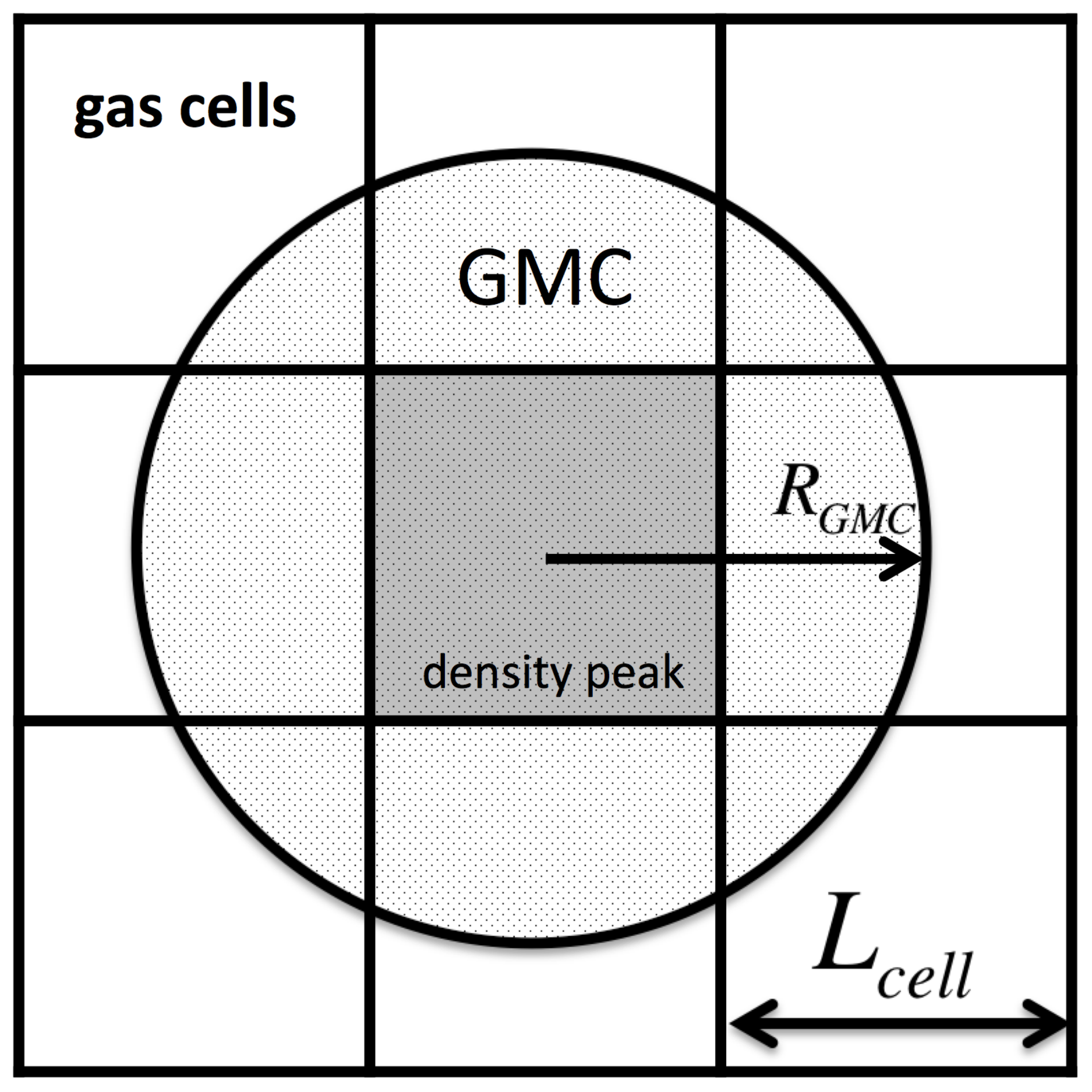}
}
  \vspace{0cm}
\caption{\small Sketch of the star-forming GMC sphere laid out on the gas cell structure in the simulation.
}\label{fig:sphere}
\end{figure}

\subsubsection{Restriction on creation of new clusters near existing active clusters}

In Paper~I, cluster particles were seeded in cold dense cells that contain local density peaks. We follow similar procedure here but with additional algorithmic improvements and constraints.

As we discussed in Section~\ref{sec:refinement}, the physical size of gas cells at the finest refinement level is always kept in the range $3-6\,$pc. This means that clusters in the smallest cells can accrete gas from all 27 cells that overlap with the GMC sphere (see Figure~\ref{fig:sphere}). To be consistent with our interpretation that the peak-density cell represents the star-forming part of a single GMC, the neighboring cells should not produce new clusters that would compete for gas supply with an existing cluster. Therefore, we prohibit a cell from creating a new particle if it already has an actively growing cluster in a neighbor cell.

This restriction makes sense only for the finest refinement level cells. If cluster particles are formed in coarser cells, which enclose a large fraction of volume of the GMC sphere, then there is no competition with neighboring cells. In this case, we allow new clusters to be created.

We also do not apply this restriction to the finest-level cells that are separated by more than one cell size from the central cell. These are the diagonal cells in the cube. Thus, the only cells that are prohibited from creating new clusters are the six cells ``face-touching'' the central cell.

To describe the above criteria for cluster seeding more quantitatively, we define an overlap fraction, $f_{\rm over}$, of the volume of a given ``face-touching'' neighbor that is occupied by the central GMC sphere. If a given cell satisfies the star formation criterion but has any ``face-touching'' neighbor with a significant overlap, $f_{\rm over} > 20\%$, that already contains an actively growing cluster, the above cell is not allowed to create a new particle. If the overlap fraction is small, $f_{\rm over} < 20\%$, we do not apply this restriction.

All of these checks are meant to minimize the impact of the neighbor restriction as much as is reasonable. However, we show in Section~\ref{sec:results} that it still has a significant impact on the shape of the CIMF, relative to the results in Paper~I.

\subsubsection{No removal of low-mass clusters} \label{sec:noremoval}

In the Paper~I algorithm, any inactive cluster particles less massive than the threshold mass of $10^3\Msun$ were recycled to speed up the simulations. Recycling meant that the material converted into stars was returned to the ISM and available for future star formation. However, some amount of energy and momentum from the stars of these failed clusters was deposited into the surrounding gas while the clusters were forming but their final mass was unknown, and this feedback could not be undone. That is, failed clusters produced some feedback that was not real. The contribution of these low-mass clusters to the stellar mass budget and overall galaxy dynamics was negligible, and therefore we can expect the impact of their feedback to be small.

With a stronger stellar feedback implementation described below in Section~\ref{sec:feedback}, both the galaxy stellar mass and the number of cluster particles are significantly reduced. Therefore, the recycling of low-mass clusters is not needed. In the new runs in this paper, we have eliminated it.

We also boosted $M_{\rm th}$ from $10^3\Msun$ to $6\times 10^3\Msun$, because the effective cluster formation time is much shorter than the maximum time $\taumax$ that is used to predict the mass of a cluster. Although the effective timescale for individual clusters varies, we show in Section~\ref{sec:result-tau} that $\sim$ 2.5~Myr is a good upper limit of the formation timescale for clusters less massive than $\sim 10^5\Msun$. Therefore, the minimum formation rate $\dot{M}_{\rm min} = 10^3\Msun/15\,$Myr in Paper I produced clusters with typical mass $\sim \dot{M}_{\rm min} \times 2.5\,\mathrm{Myr}$, a factor of 6 below our intended threshold. With the new value $M_{\rm th} = 6\times 10^3\Msun$ we eliminate most of small clusters with final masses below $10^3\Msun$.

\subsubsection{No molecular fraction threshold for continuing cluster growth}\label{sec:molecular-threshold}

In Paper~I, we employed a threshold on the molecular fraction of hydrogen so that clusters could only form and grow in cells with a high molecular fraction, $f_{\rm H_2}>50\%$. This meant that if the molecular fraction of a given cell fell below $50\%$, not only were new clusters not allowed to form but also an existing active cluster in the cell was not allowed to continue to grow. 

After detailed analysis of the growth history of individual cluster particles, we found that this molecular fraction threshold leads to intermittent gas accretion for some clusters, when $f_{\rm H_2}$ oscillates around the threshold value. It also generates many gap periods without any mass growth, especially for the most massive clusters. These gaps affect the calculation and interpretation of the cluster formation timescales. Indeed, we found that the long formation timescales of some massive clusters in Paper~I were not due to slow star formation but due to the cessation of star formation by this molecular threshold.

Here we revise the implementation of the molecular threshold by adopting it only as the criterion for initial particle creation. Once a cluster is formed, its subsequent accretion is no longer affected by the molecular threshold. Thus, active clusters can continue to grow their mass at the rate that is proportional to the molecular fraction, via Equation~(\ref{eq:sfr}), until the gas density drops below $n_{\rm crit} = 10^3\, {\rm cm}^{-3}$.

\subsubsection{Gap elimination} \label{sec:gap-elim}

Another reason for the existence of long accretion gaps is due to the motion of clusters. In high-density regions, such as the inner part of the galaxy, gravitational forces accelerate particles to high velocities. For a velocity dispersion on the order of 20~km~s$^{-1}$, during the active formation period, clusters can travel up to 300~pc, which is much larger than the cell size or the GMC size. This means that active clusters can travel through and accrete gas from multiple GMCs, different from the one in which they had originally formed. This is not physically correct, because a cluster should accrete gas from only one GMC, and each GMC should host a distinct cluster. Moreover, when a cluster travels through low-density regions between dense clouds, the accretion stops because the gas density falls below the threshold density for star formation, thus creating artificial gaps in the mass growth history.

To eliminate accretion of a given cluster particle from multiple GMCs, in the simulations presented in this paper, we monitor the accretion process for each active cluster. We deactivate a cluster even before it reaches the maximum time $\taumax$ if we find that it completely stopped growing mass for more than 1~Myr. By design, this procedure eliminates all gaps longer than 1~Myr, and systematically reduces the cluster formation times even for the most massive clusters, as we show in Section~\ref{sec:result-tau}.

Conversely, any effect that increases the relative velocity of an actively growing cluster and its parent cloud would shorten the formation timescale. Strong momentum feedback into an inhomogeneous medium may dislodge the clusters and terminate their growth, even before exhausting the gas supply.

\subsubsection{Mass-loss rate due to stellar evolution} \label{sec:massloss}

In Paper I, a stellar particle lost mass due to passive stellar evolution and stellar winds at the rate given by
\begin{equation} \label{eq:massloss-ART}
  \frac{dM}{dt} = \frac{\eta}{t+\tau_{\rm loss}}\, M_{t=0},
\end{equation}
where $\eta$ is the typical fraction of mass loss and $\tau_{\rm loss}$ is the characteristic timescale, both of which depend on the IMF of stars. For a \citet{kroupa01} IMF, we used $\eta=0.046$ and $\tau_{\rm loss} = 2.76\times 10^{5}\,$yr. The time evolution of the cumulative fraction of stellar mass loss is then obtained by integrating Equation~(\ref{eq:massloss-ART}) over time:
\begin{equation}
  f_{\rm loss}(t) = \frac{M_{\rm loss}}{M_{t=0}}=\int_0^t \frac{\eta}{t+\tau_{\rm loss}}dt 
  = \eta\ln{\left(\frac{t}{\tau_{\rm loss}}+1\right)}.
\end{equation}

\begin{figure}[t]
\includegraphics[width=1.0\hsize]{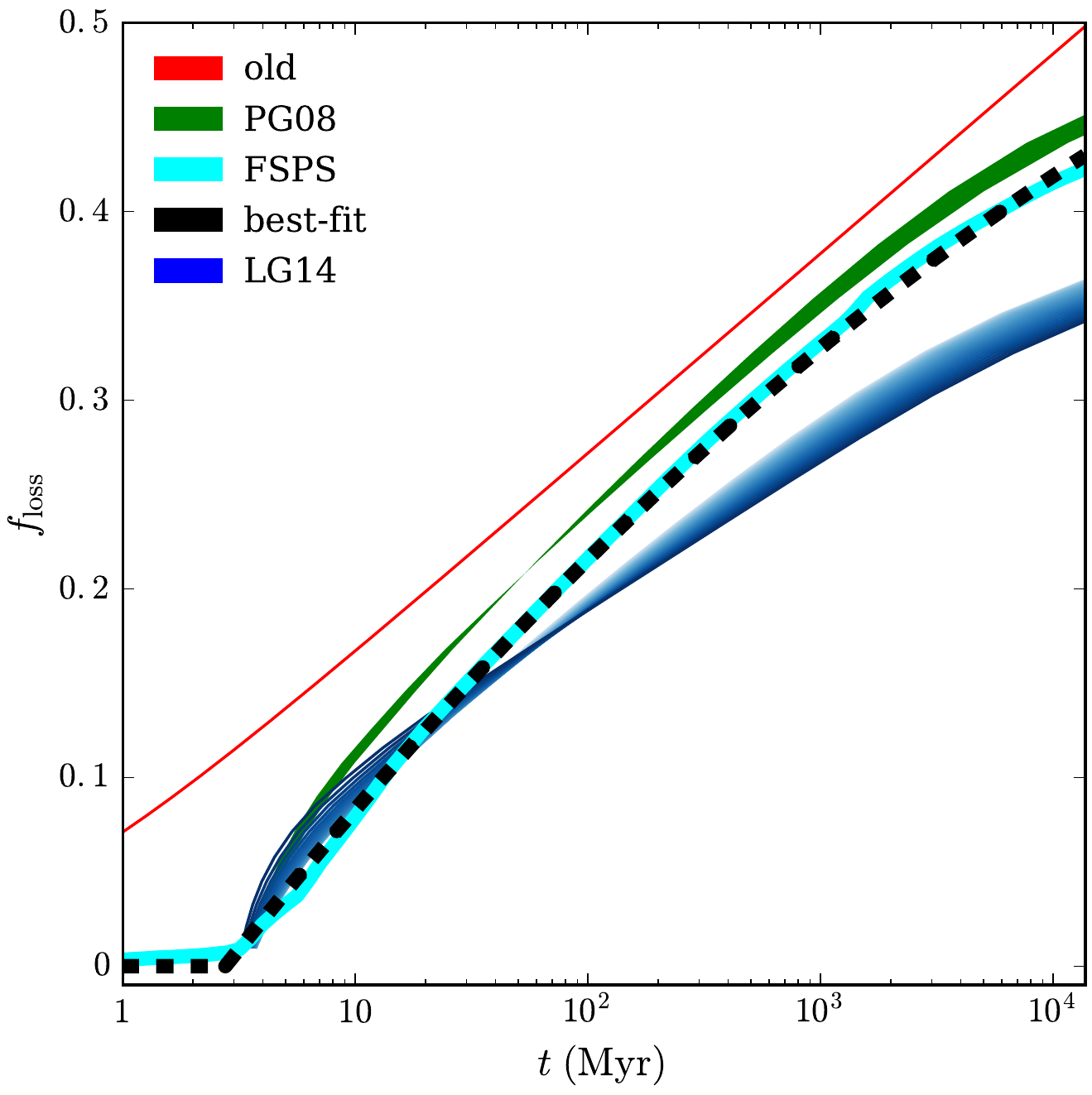}
\caption{\small Cumulative fraction of mass loss due to stellar evolution for a single stellar population with a \citet{kroupa01} IMF: old prescription used in Paper~I (red), \citet{prieto_gnedin08} (green), \citet{li_gnedin14} (blue), and FSPS (cyan). The line width for each source represents the range of mass loss from stars of different metallicity, from zero to solar. The best-fit expression to the average FSPS result (described by Equation~\ref{eq:massloss-new}) is overplotted by the thick dashed line.
}\label{fig:massloss}
\vspace{1mm}
\end{figure}

Figure~\ref{fig:massloss} shows this mass-loss history, together with the results from the detailed stellar population synthesis model FSPS\footnote{Flexible stellar population synthesis, \url{https://github.com/cconroy20/fsps}} for the same IMF \citep{conroy_etal09,conroy_gunn10}. We find that the previous implementation overestimates the mass-loss rate over the whole stellar lifetime. The overestimate is more prominent for younger stellar populations. For example, after 3~Myr about 10\% of the stellar mass is lost in Paper I modeling but less than 1\% in FSPS modeling.

Here we introduce a new mass-loss model that fits the time evolution of the cumulative fraction in FSPS with a second-order polynomial. We write $f_{\rm loss}(t) = ax^2+bx+c$, where $x=\log_{10}(t/\rm yr)$, and obtain the best-fit parameters: $a=-0.010$, $b=0.288$, and $c=-1.42$. The mass-loss rate at a given age $t$ is obtained by differentiating $f_{\rm loss}(t)$ with respect to $t$:
\begin{equation} \label{eq:massloss-new}
  \frac{dM}{dt} = \frac{b + 2a \log_{10}(t)}{\ln 10} \, \frac{M_{t=0}}{t}.
\end{equation}
Note that this expression works for stars in the age range from 2.75~Myr to 13.7~Gyr. For ages younger than 2.75~Myr, we assume no appreciable mass loss.

\subsubsection{Changes to stellar feedback} \label{sec:feedback}

As in Paper~I, the feedback subgrid model consists of mass, momentum, and energy injections from stellar winds, radiation pressure, and SN explosions. The main update in this paper is in the implementation of SNRs feedback. In the new SNR model, we estimate the partition of the thermal, kinetic, and turbulence energies of SNR using a parameterization model calibrated by the high-resolution hydrodynamic simulations in an inhomogeneous turbulent medium by \citet{martizzi_etal15}. The energy and momentum input from SNRs depends on the ambient density and spatial resolution, as described in detail in \citet{semenov_etal16}.

One caveat of the SNR model is that it is calibrated by the simulations of SN explosion in isolation, rather than in a more complex star forming environment. In reality, clusters produce a large number of massive stars that undergo SN explosion over several Myr. \citet{gentry_etal17} found that such clustering of SNe can enhance momentum feedback by an order of magnitude relative to that delivered by an isolated SN. Another important motivation of boosting the injected momentum is to compensate for the momentum loss due to advection errors as the SN shell moves across the simulation grid \citep[e.g.][]{semenov_etal17}. Therefore, in our new runs, we boost the momentum feedback of the \citet{martizzi_etal15} SNR model by a factor $\fboost=$3, 5, or 10, as listed in Table~\ref{tab:run-parameter}. 

\subsubsection{Initial bound fraction} \label{sec:initial_bound}

In Paper I, we assumed that the whole accreted mass of cluster particles during their active growth is gravitationally self-bound when the particles emerge from their natal clouds. That is, the stellar particle mass is the bound cluster mass. This assumption does not take into account the complex dynamical evolution of star clusters in the early phase, when the boundedness is affected by the hierarchical structure of the ISM and gas expulsion due to stellar feedback. Recent observations and numerical simulations of turbulent clouds suggest that the fraction of mass in a given star-forming complex that remains bound after a few Myr depends strongly on the star formation efficiency on the scale of GMCs \citep{goodwin97, geyer_burkert01, goodwin_bastian06, smith_etal11, kruijssen_etal12}.

In this paper we introduce the \textit{initial bound fraction}, $f_i$, and assign it to each stellar particle. We adopt a linear dependence of $f_i$ on the local star formation efficiency,
\begin{equation} \label{eq:initial_bound}
  f_i \equiv \min{\left(\frac{\epsilon_{\rm int}}{\epsilon_{\rm core}}, 1\right)}
\end{equation}
where $\epsilon_{\rm core} = 0.5$ is the correction factor for mass loss due to protostellar outflows, suggested by \citet{kruijssen12}, and $\epsilon_{\rm int}$ is the integral star formation efficiency, defined as the ratio between the full mass of the active stellar particle (here we write it explicitly as $M_{\rm f}$, but later omit the subscript "f") and the maximum baryon mass of the GMC throughout the whole course of cluster accretion: 
\begin{equation}
  \epsilon_{\rm int} \equiv \frac{M_{\rm f}}{\max_{\rm time}(M_{\rm star}+M_{\rm gas})}.
  \label{eq:epsint}
\end{equation}
Note that the linear relation between local star formation efficiency and initial bound fraction is based on the analysis of the star formation simulations of \citet{bonnell_etal08}. Since those simulations only cover high-efficiency cases with $f_i>0.4$, it is unclear whether such linearity is valid for low-efficiency GMCs. Recent simulations of isolated GMCs \citep{goodwin09, dale_etal14, gavagnin_etal17} suggest that the bound fraction depends not only on $\epsint$ but also on the dynamical state of the young stars before gas expulsion by feedback. Including this effect in future work could improve the model.

It is important to emphasize that the GMC mass in our simulations varies on a timescale of less than a Myr, as the central part is converted into stars and outside gas flows in, in essentially free fall. The sum of stellar and gas mass usually increases over several Myr before declining as the feedback of young stars disperses the remaining gas. This situation is very different from simulations of star formation in isolated clouds, where the total baryon mass is fixed as the initial cloud mass. 

Bound cluster mass in our model is then defined as the stellar particle mass at the end of the active growth period times the initial bound fraction: $M = f_i \, M_{\rm f}$. The fraction of stars remaining bound to the cluster continues to evolve due to the dynamical evaporation and tidal stripping. We calculate this process with a subgrid model and will discuss the evolution of the cluster mass function in a follow-up paper. We attach the initial bound fraction variable to every cluster particle and assume that the unbound part remains near the cluster, so that both the bound and unbound parts move together under the same gravitational potential.

\subsubsection{Alternative definition of cluster formation timescale} \label{sec:updates-tau}

Although the actual mass growth history of individual star clusters is difficult to determine in observations, recent theoretical models and simulations suggest a dynamic, time-variable process. Self-gravitating collapse of the dense parts of GMCs is thought to accelerate star formation until stellar feedback, in the form of stellar winds and radiative pressure/heating, pushes the accreting gas out. The whole process happens quickly, as indicated by the observed age spread of stars in embedded clusters within 3--4~Myr, only a couple freefall times of their natal clouds; see Section~\ref{sec:cft-obs}.

One way to quantify the age spread is to estimate the timescale for the formation of a given fraction of stars within a cluster, e.g. the period when the cluster mass grows from 10\% to 90\% of its final mass. However, since the final mass of the clusters in our model is not known until their growth is terminated by the feedback, obtaining the 10-90\% timescale requires tracking the detailed growth history of all clusters, which is computationally prohibitive. Therefore, in Paper I, the average duration of cluster formation was calculated as the mass-weighted timescale,
\begin{equation} \label{eq:timescale}
  \tauave \equiv \frac{\int_{0}^{\taumax} t \, \dot{M}(t) \, dt}{\int_{0}^{\taumax} \dot{M}(t) \, dt},
\end{equation}
where $\dot{M}(t)$ is the cluster SFR at time $t$. This definition best describes steady mass accretion. For example, in the case of constant SFR, $\tauave = \taumax/2$. 

One caveat of this definition is that, in the case of increasing and then decreasing SFR, $\tauave$ actually records the epoch when $\dot{M}$ reaches its peak, rather than the duration of the process. Here we introduce a new quantity that better describes the width of such SFH. The age spread is defined as the ratio between the full particle mass $M_{\rm f}$ and the mass-weighted SFR over the whole growth history $\langle\dot{M}\rangle$:
\begin{equation}
  \tauspread \equiv \frac{M_{\rm f}}{\langle\dot{M}\rangle}
  = \frac{M_{\rm f}}{\int_0^{\taumax} \dot{M}^2 dt/M_{\rm f}}.
\end{equation}

\begin{figure}[t]
\includegraphics[width=1.0\hsize]{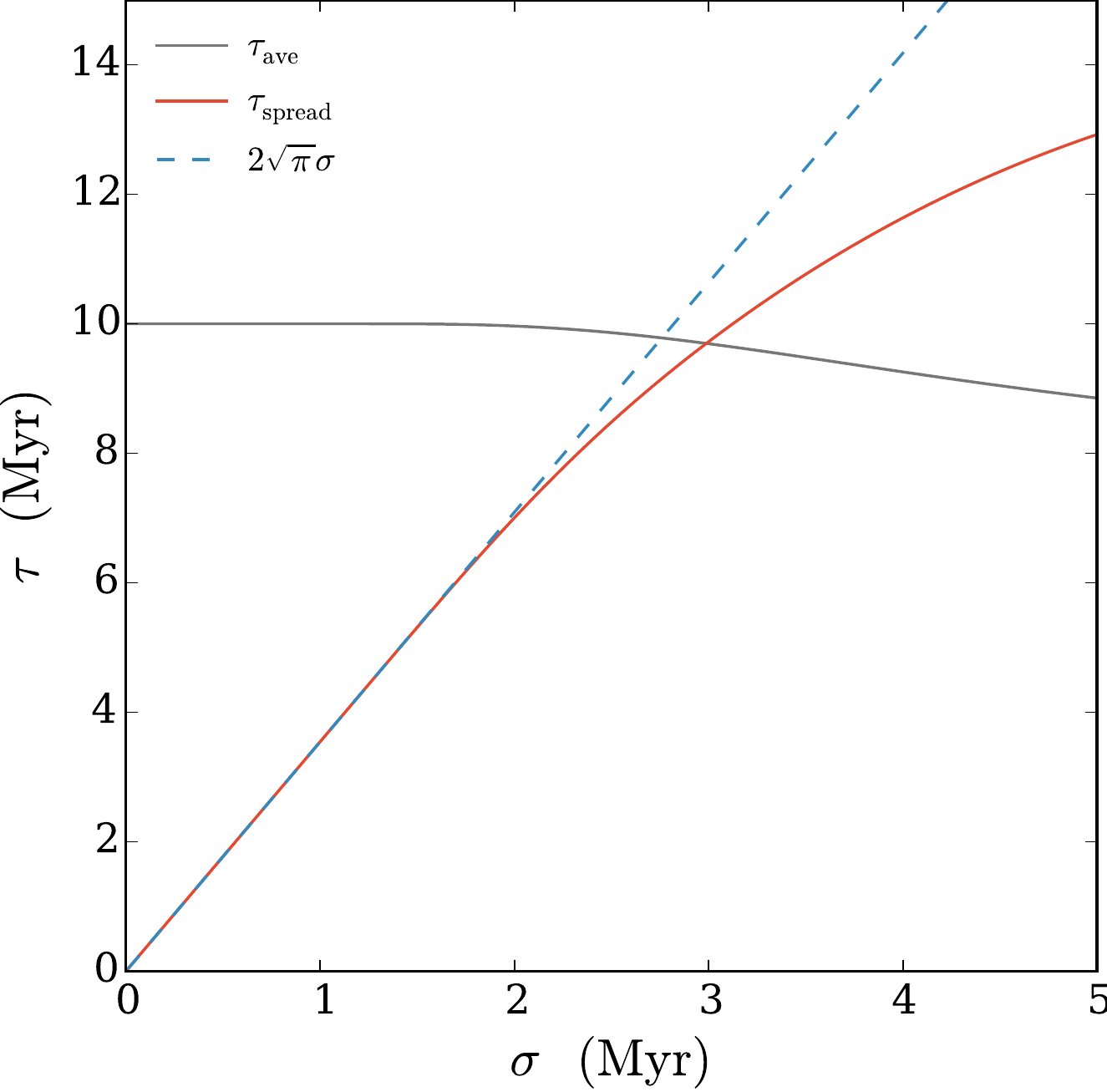}
\caption{\small Example of the old ($\tauave$, black) and new ($\tauspread$, red) definitions of cluster formation timescale as a function of Gaussian width of the SFR, peaked at a relatively late time $t_0=10$~Myr. The new definition closely follows the width of the accretion rate when $\sigma$ is small compared to the total duration of the star formation episode.
}\label{fig:tau-sigma}
\vspace{1mm}
\end{figure}

For a power-law mass accretion history, $\dot{M}\propto t^{\alpha}$, $\alpha\ne-1$, both $\tauave$ and $\tauspread$ can be explicitly evaluated as 
\begin{equation}
  \tauave = \frac{\alpha+1}{\alpha+2} \taumax \quad\mathrm{and}\quad \tauspread = \frac{2\alpha+1}{(\alpha+1)^2} \taumax.
\end{equation}
However, in the case when $\dot{M}$ exhibits a peak, e.g. a Gaussian function $\dot{M} \propto \exp{(-(t-t_0)^2/2\sigma^2)}$,
\begin{equation}
  \tauave = t_0 + \sqrt{\frac{2}{\pi}}\sigma\frac{\exp{(-t_0^2/2\sigma^2)-\exp{(-(t_0-\taumax)^2/2\sigma^2)}}}{\erf{(t_0/\sqrt{2}\sigma)}+\erf{((\taumax-t_0)/\sqrt{2}\sigma)}},
\end{equation}
while the new definition gives
\begin{equation}
  \tauspread = \sqrt{\pi}\sigma \frac{[\erf{(t_0/\sqrt{2}\sigma)}+\erf{((\taumax-t_0)/\sqrt{2}\sigma)}]^2}{\erf{(t_0/\sigma)}+\erf{((\taumax-t_0)/\sigma)}}.
\end{equation}
The relationship between the intrinsic width of the mass accretion history $\sigma$ and the derived ages $\tauave$ and $\tauspread$ is shown in Figure~\ref{fig:tau-sigma}.

When $\sigma \to 0$, the SFH reduces to a $\delta$-function located at $t=t_0$. In this case, $\tauave \to t_0$, while $\tauspread \to 2\sqrt{\pi}\sigma$. This extreme case illustrates the problem with our previous definition. Rather than recording the age spread of stars, $\tauave$ actually reflects the epoch of the peak of star formation. Instead, the new definition follows the true width of the age distribution.

As $\sigma$ increases, the SFH becomes flatter. At the other extreme, if $\sigma > \taumax$, the SFR can be considered constant. The above equations then reduce to $\tauave \to t_0/2$ and $\tauspread \to t_0$.

\begin{table*}
\centering
\caption{Model runs}
\begin{tabular}{lcclc}
\tableline\\[-2mm]
Name 		& $z_f$	& $\epsff$ & Feedback & $L_{\rm cell}$ (pc) at $z=2$ \\[1mm]
\tableline\\[-2mm]
SFE10 		& 1.7	& 0.1      & early$^a$+5*SNR$^b$ & 5 \\
SFE50 		& 1.5	& 0.5      & early+5*SNR         & 5 \\
SFE50-SNR3 	& 2.0	& 0.5      & early+3*SNR         & 5 \\
SFE100	 	& 1.5	& 1.0      & early+5*SNR         & 5 \\
SFE200	 	& 1.5	& 2.0      & early+5*SNR         & 5 \\
SFEturb 	& 2.7	& Variable & early+5*SNR         & 5 \\
\tableline\\[-2mm]
fid-PaperI 	& 3.2	& 0.1      & early+old SN$^c$	 & 20 \\
SNR10-old 	& 2.0	& 0.1      & early+10*SNR        & 10 \\
\tableline\\
\end{tabular}\\
Notes: $a.$ "early": early feedback schemes including stellar wind and radiative pressure. $b.$ "SNR": resolution-dependent SNR feedback scheme in \citet{martizzi_etal15}; the number before "SNR" is the boosting factor. $c.$ "old SN": previous SN thermal and momentum feedback with the amount calibrated by \citet{agertz_kravtsov_15}. 
 \label{tab:run-parameter}
\end{table*}

\begin{table*}
\centering
\caption{Compilation of the observed properties of nearby young star clusters} \label{tab:age-spread}
\begin{tabular}{lcccl}
\tableline\\[-2mm]
Name		&	Age (Myr)	& Age Spread (Myr)	& Mass ($\Msun$) &  Reference \\
\tableline\\[-2mm]
Orion Nebula Cluster & 4-5 & 1-3 & 2000	& \citet{dario_etal10a,jeffries_etal11} \\
R136		 		& 4-5 &	3 &	4.5e5 & \citet{massey_hunter98} \\
LH95				& 4	& 2.8-4.4 & ? & \citet{dario_etal10b} \\
NGC 1569A			& $<5$ & free of dust extinction & 1e6 & \citet{maoz_etal01} \\
Westerlund 1 		& 5	& 0.4 or $<1$ & 6.3e4 & \citet{negueruela_etal10,kudryavtseva_etal12} \\
NGC 4103			& ?	& 2-4 & ? & \citet{forbes96} \\
NGC 3603 YC			& 2	& 0.1 & ? & \citet{kudryavtseva_etal12} \\
NGC 3603 HII		& 1 & $<1$ & 1.9e4 & \citet{pang_etal13} \\
W3 Main				& ?	& 2-3 & ? & \citet{bik_etal12} \\
Antennae clusters	& ?	& $<6, A_V=1$ mag & ? & \citet{whitmore_zhang02} \\
NGC 4449 clusters	& ?	& $<5, A_V=0.5-1.5$ mag & 0.5-5e4 & \citet{reines_etal08} \\
M83	clusters		& ?	& $<4$ & ? & \citet{hollyhead_etal15} \\
\tableline\\[-1.8mm]
\end{tabular}
\end{table*}

\subsection{New runs}

In this paper we present new runs with the above improvements. All start with the same initial conditions as in Paper~I. Their key physical parameters are listed in Table~\ref{tab:run-parameter}. The number after ``SFE'' in their names corresponds to the local $\epsff$ in percent. One exception is the ``SFEturb'' run, with variable turbulence-dependent $\epsff$ based on the parameterization of $\epsff$ as a function of cell turbulence energy \citep{padoan_etal12}. The evolution of the turbulence energy is calculated by a subgrid-scale model from \citet{schmidt14}, implemented in the ART code by \citet{semenov_etal16}. The standard value of the SNR momentum boost factor is $\fboost=5$. In run SFE50-SNR3, we adopt $\fboost=3$ to test the dependence of global SFR on this factor.

In addition to these new runs with full updates, we include also for comparison the fiducial run "old-fid" from Paper I, and the "SNR10-old" run with $\fboost=10$ and some intermediate degree of update. These latter runs illustrate progression in the development of our algorithm.

$\bullet$ fid-PaperI:
The fiducial run in Paper I with $\epsff=10\%$. As shown in Figure~\ref{fig:sfh}, this run overestimates the expected SFR of the main galaxy by more than one order of magnitude at $z=4-8$. Due to the high SFR, a large number of clusters more massive than $10^{5}\Msun$ are formed at high $z$. The inability to suppress SFR at high $z$ is caused by the weak feedback that is used in this run. Another sign of the ineffectiveness of the feedback comes from the mass-weighted cluster formation timescale $\tauave$. In Paper~I, we showed that the median value of $\tauave$ for all clusters in this run is about 3~Myr, consistent with the observational constraint that is described in Section~\ref{sec:cft-obs}. Detailed analysis suggests that this median value is dominated by a large number of less massive cluster particles. For clusters more massive than $10^{5}\Msun$, however, $\tauave$ is much longer, sometimes even longer than $\taumax/2 = 7.5\,$Myr, suggesting that the stellar feedback cannot terminate the gas accretion process for massive clusters within the allowed accretion timescale that is assigned in the simulation.

$\bullet$ SFE10-old:
Same as ``fid-PaperI'', except (1) a refinement strategy that keeps the ratio of $\Rgmc/L_{\rm cell}$ roughly constant but one level coarser than that described in Section~\ref{sec:refinement}; and (2) a new SNR feedback prescription described in Section~\ref{sec:feedback} with a momentum boosting factor $\fboost=5$. We found that the SFH of this run follows the abundance matching results until $z\approx 2.5$, when a major merger happens in the main galaxy. This merger brings so much cold gas into the inner regions of the galaxy that stellar feedback is unable to disperse it. An intense starburst occurs, and a prominent stellar spheroid forms at the galactic center. As we discuss in the next section, adding one additional refinement level prevents this starburst and leads to a reasonable SFH. Another result is that, although the SFR of this run is much smaller than that of ``fid-PaperI'' due to the adoption of the new feedback model, there are still many clusters, especially massive ones, that have very long formation timescales. This finding inspired a detailed investigation of the origin of the long timescale and the implementation of gap elimination algorithms, which we describe in Sections~\ref{sec:molecular-threshold} and \ref{sec:gap-elim}.

$\bullet$ SNR10-old:
Same as ``SFE10-old'', but with a larger boosting factor $\fboost=10$. The goal of this run is to probe the upper limit of $\fboost$. As shown in Figure~\ref{fig:sfh}, the SFH is always under the abundance matching result. This underestimate of SFR sets an upper limit on the reasonable choice of $\fboost$.

For comparison, we have also run a simulation with the standard $\fboost=5$ but boosted the momentum feedback from radiative pressure by a factor of 10. This boost produced no apparent effect on SFH, unlike the SN momentum boost.

\begin{figure*}[t]
\includegraphics[width=1.0\hsize]{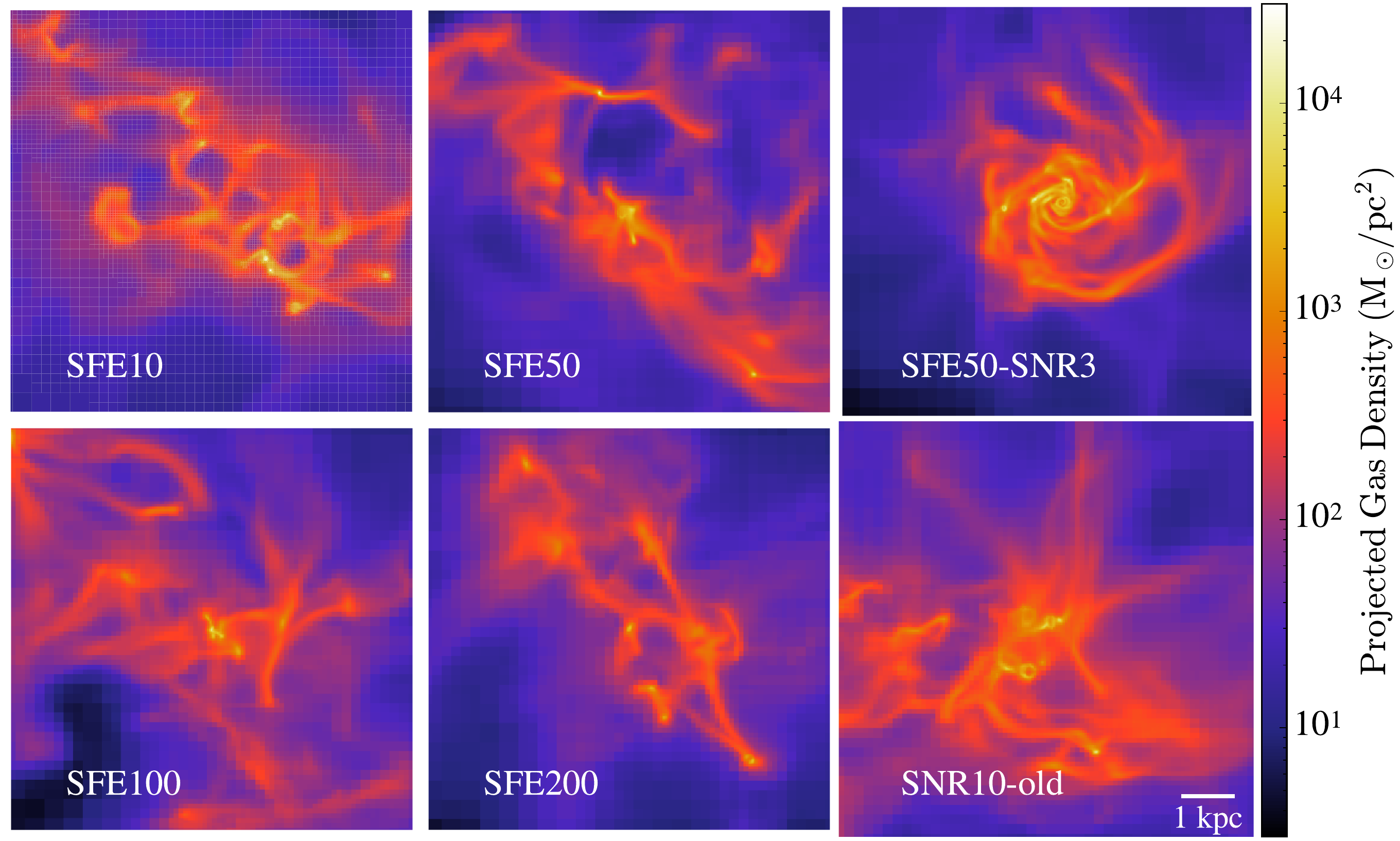}
\caption{\small Gas density projection plots of the main galaxy at $z\approx 2$ for different runs. The adaptive refinement structure of the oct-tree code is shown in the upper left panel, and the length scale of 1~kpc is shown in the lower right panel.
}\label{fig:gas-proj}
\end{figure*}

\subsection{New observational constraints}
\subsubsection{Observed age spread of young clusters} \label{sec:cft-obs}

To use the predicted formation duration as an additional test of the models, in Table~\ref{tab:age-spread}\ we compile recent measurements of the spread of the relative ages of stars in several young star clusters. Although obtaining accurate age measurements is still challenging, current observations suggest that the age spread should be less than 6~Myr for various star-forming regions in different galaxies. This upper limit also agrees with small-scale hydrodynamic simulations of cluster formation, which suggest that the star formation process should proceed on a timescale comparable to the freefall time \citep[e.g.][]{hartmann_etal12, grudic_etal18}. This relatively short timescale provides a strong constraint on the implementation of star formation and stellar feedback.

\subsubsection{SFR vs. maximum cluster mass in nearby galaxies} \label{sec:Mmax-obs}

The observed data of SFR surface density and V-band absolute magnitudes of the brightest young clusters in different galaxies is compiled in \citet{adamo_etal15}. Note that this compilation contains observations of galaxy-wide measurements, as well as spatially resolved samples. To convert the magnitudes of the brightest clusters to the corresponding stellar masses, we used the V-band mass-to-light ratio for young star clusters from \citet{lieberz_kroupa17}: $M/L_{V} = 0.014\Msun/L_{V\odot}$.

\section{Results} \label{sec:results}

Most of the simulations presented here were performed at the high-performance computing center Flux at the University of Michigan. We highlight here that, although our simulations incorporate many state-of-the-art physical processes, such as non-equilibrium chemical networks and radiative transfer, with very high spatial resolution, the total computing time is not huge. For reference, runs with $\epsff \geqslant 0.5$ take about $30,000-50,000$ CPU hr to reach $z\approx 1.5$.

Figure~\ref{fig:gas-proj} shows the gas surface density of the inner 4~kpc of the main galaxy at $z\approx 2$ for six different runs. The projection is taken along the X-axis of the simulation box, which is close to the intrinsic rotation axis of the disk. Compared to the fiducial run in Paper~I, most of the density maps here do not exhibit well-defined gaseous disks. Instead, due to the stronger feedback implementation, the most prominent structures are the kpc-scale low-density cavities surrounded by rings of higher-density beads and filaments. These cavities are created by multiple SNe from young star clusters. Shock waves of the ``superbubbles'' can travel several kpc through these low-density regions, compress the gas located around the edge of the bubble, and possibly trigger subsequent star formation. They also generate large-scale outflows from the inner regions. The densest regions of the galaxies, therefore, are not centrally concentrated but are distributed as many kpc-scale clumps, which is consistent with recent Hubble Ultra Deep Field observations of $z\approx2$ star-forming galaxies \citep{guo_etal12}. 

On the other hand, the SFE50-SNR3 run with weaker feedback (upper right panel) shows a more regular disk with centrally-peaked gas distribution and prominent spiral arms. This contrast illustrates that the large-scale gas distribution is very sensitive to variation of the momentum feedback parameters, even by a factor of two.

\begin{figure*}
\centerline{
\includegraphics[width=0.48\hsize]{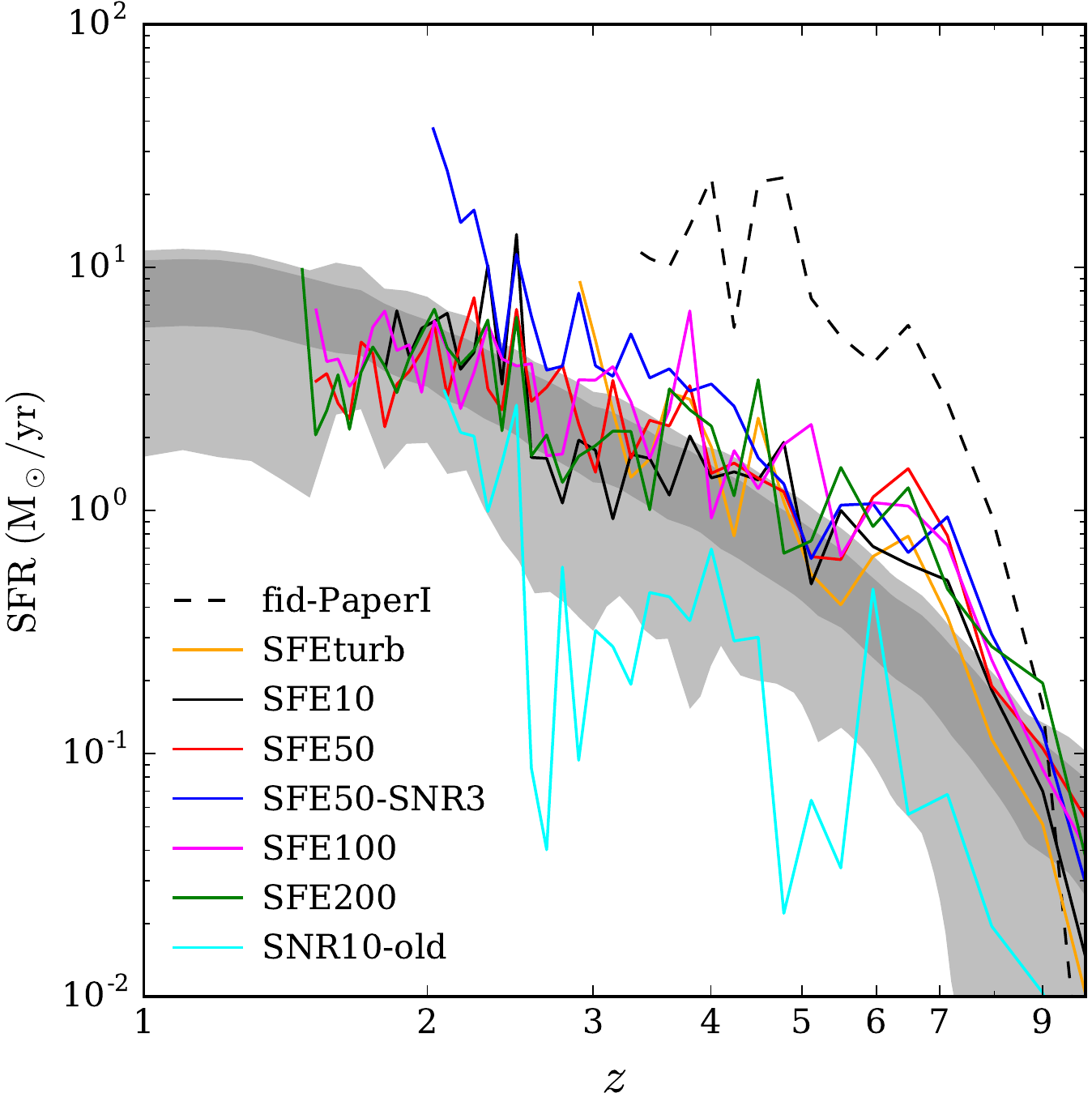}
\hspace{0.01\hsize}
\includegraphics[width=0.5\hsize]{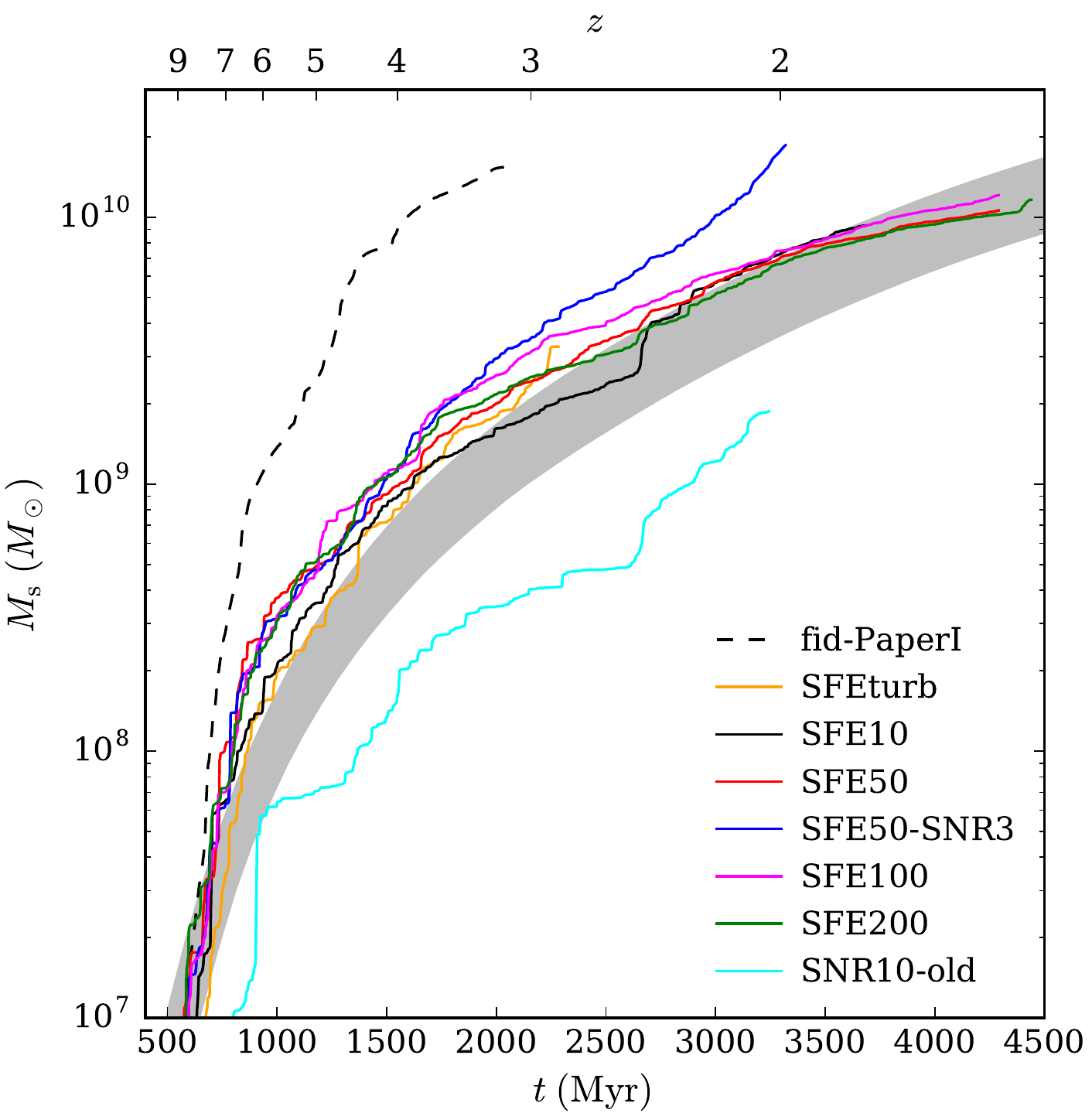}
}
\caption{\small {\it Left:} SFH of the main galaxy for runs with different star formation and feedback parameters. The SFR is derived from all stellar particles within the main galaxy at the last snapshot, see Section~\ref{sec:sfh} for details. Dark and light shaded areas are 1-$\sigma$ and 2-$\sigma$ confidence intervals of the expected SFR for an average $10^{12}\Msun$ halo from abundance matching \citep{behroozi_etal13a}. {\it Right:} cumulative stellar mass history of the main galaxy. The shaded area is the $1\sigma$ confidence interval from the abundance matching result.
}\label{fig:sfh}
\end{figure*}

\subsection{SFH of the main galaxy} \label{sec:sfh}

In Figure~\ref{fig:sfh}, we show the SFH of the main halo for several runs with different values of the star formation and feedback parameters. The SFH is calculated from all cluster particles within $0.5\, R_{\rm vir}$ of the main galaxy from the last available snapshot of each run, where $R_{\rm vir}$ is the virial radius of the dark matter halo. The rate of star formation is averaged over 100~Myr around a given epoch in order to smooth out stochasticity.

In general, simulations with $\fboost=5$ agree well with the abundance matching results, in the sense of both SFR and the cumulative stellar mass growth. This suggests that the strength of stellar feedback used in these runs is appropriate to reproduce the inefficient star formation at high redshift. Changing the intensity of feedback has a measurable effect on the SFH. The run with $\fboost=3$ systematically overestimates SFR at all times, while the run with $\fboost=10$ underestimates SFR by an order of magnitude. The resulting SFHs of the two runs are well beyond the $1\sigma$ confidence intervals of the abundance matching result. Therefore, the choice of $\fboost$ in our simulations is bracketed between 3 and 10.

On the other hand, varying $\epsff$ from $\sim 0.01$ to 2.0 has almost no systematic effect on SFH. This insensitivity to the value of $\epsff$ has already been found in several recent cosmological simulations with the implementation of strong stellar feedback and high spatial resolution \citep{agertz_etal13, hopkins_etal13, agertz_kravtsov_15}. The feedback-controlled low-efficiency star formation activity can be interpreted as the short lifetime of star-forming regions, which then requires a large number of star-forming cycles to convert all the gas into stars \citep{semenov_etal17}.

However, the lack of sensitivity of SFH to the local star formation efficiency does not mean that one can assign arbitrary values to $\epsff$ in galaxy formation simulations. As we show in the rest of this section, $\epsff$ has dramatic effects on the properties of individual star-forming regions and, in turn, the properties of young star clusters formed within.

\subsection{KSR} \label{sec:KS}

Above, we showed that with $\fboost=5$ our simulations can reproduce the SFH of Milky Way-sized galaxies, suggesting that the galaxy-wide star formation activity is well suppressed with our feedback implementation. Another commonly used global representation of star formation is the KSR, which is the relationship between the gas surface density and the surface density of the SFR. 

\begin{figure}
\includegraphics[width=1\hsize]{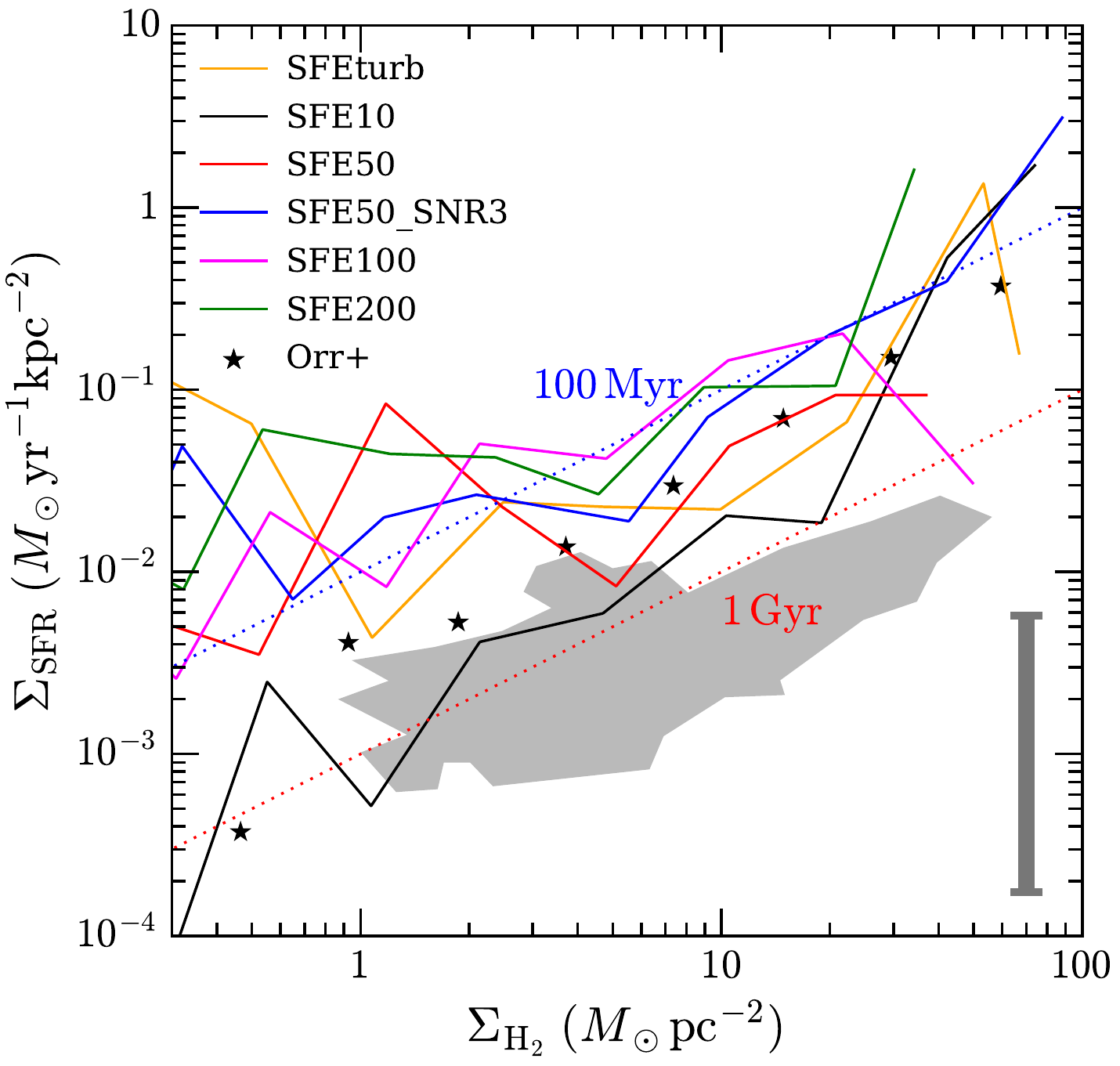}
\caption{\small The KSR for the main galaxy. The SFR is calculated over 20~Myr, spatially averaged over 1~kpc squares; see Section~\ref{sec:KS} for details. For the SFE10 to SFE200 runs, the measurements are derived from all snapshots between $z\approx9$ and $z\approx1.5$, while for the SFEturb run they are from snapshots at $z > 2.7$. The interquartile (25-75\%) range of $\Sigma_{\rm SFR}$ is shown as a thick gray bar in the lower right corner. The KSR derived from the FIRE simulations \citep{orr_etal18} on similar spatial (1~kpc) and time (10~Myr) scales is shown by black stars. The gray shaded region shows the range of observed values in nearby spiral galaxies by \citet{bigiel_etal08, bigiel_etal11}. Dotted lines indicate constant gas depletion timescales of 100~Myr (blue) and 1~Gyr (red).
}\label{fig:KS_H2}
\end{figure}

To derive the KSR for molecular gas in our simulations, we split the main galaxy disk into several $1\times1$~kpc square areas and calculate $\Sigma_{\rm H_2}$ and $\SigmaSFR$ within each area. Here $\SigmaSFR$ is estimated by using clusters younger than 20~Myr. Varying this timescale in the range of 15--50~Myr does not affect the results. We considered different spatial smoothing scales and found that using smaller scales between 100 and 500~pc produces larger scatter of the KSR but with similar median values to the 1~kpc case. We collect measurements from $z\approx 9$ to the last available snapshot in each run and calculate the median value of $\SigmaSFR$ in a given range of $\Sigma_{\rm H_2}$. Figure~\ref{fig:KS_H2} shows the KSR in six simulations with different $\epsff$. 

When expressed through molecular H$_2$ gas, the observed KSR in nearby spiral galaxies at $z=0$ is consistent with linear and does not vary significantly with metallicity or galaxy type \citep{bigiel_etal08, bigiel_etal11}. In our simulations, we do not find a systematic redshift evolution of the KSR in various snapshots from $z\approx9-1.5$. Our simulations produce an approximately linear relation but with systematically higher $\SigmaSFR$ than observed at $z=0$, by about a factor of 2-10. We also find that the simulations with higher $\epsff$ tend to have somewhat higher $\SigmaSFR$ for a given $\Sigma_{\rm H_2}$, but this trend is overwhelmed by the very large scatter.

\begin{figure*}[t]
\includegraphics[width=1.0\hsize]{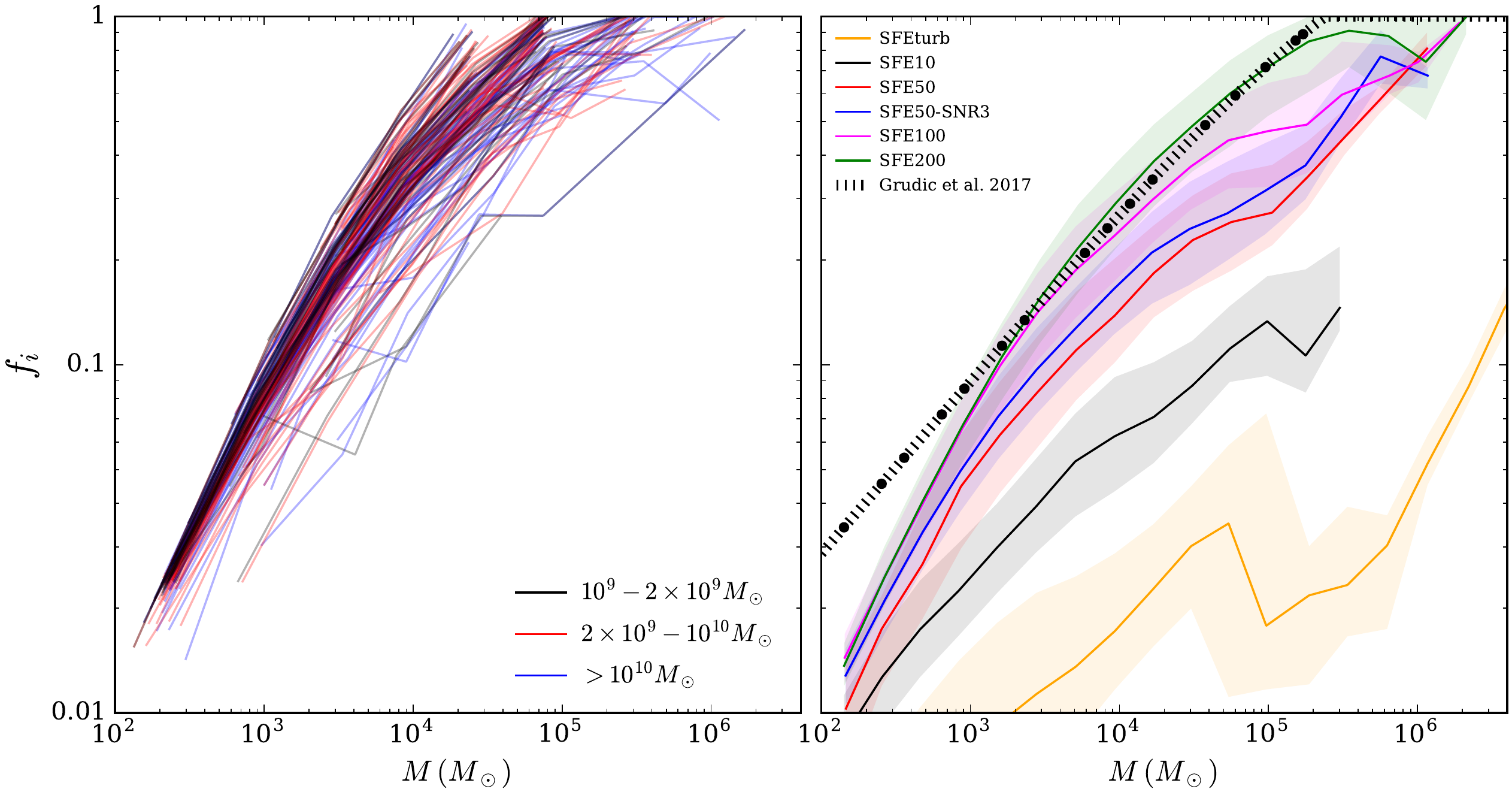}
\caption{\small {\it Left:} initial bound fraction vs. stellar particle mass in the SFE200 run. Clusters formed within galaxies of different halo mass are labeled with different colors. {\it Right:} same as left panel but for runs with different star formation and feedback parameters. The striped line shows the relation calculated from MHD simulations of star formation in isolated GMCs by \citet{grudic_etal18}.
}\label{fig:initial_bound}
  \vspace{0.3cm}
\end{figure*}

\begin{figure*}[t]
\includegraphics[width=1.0\hsize]{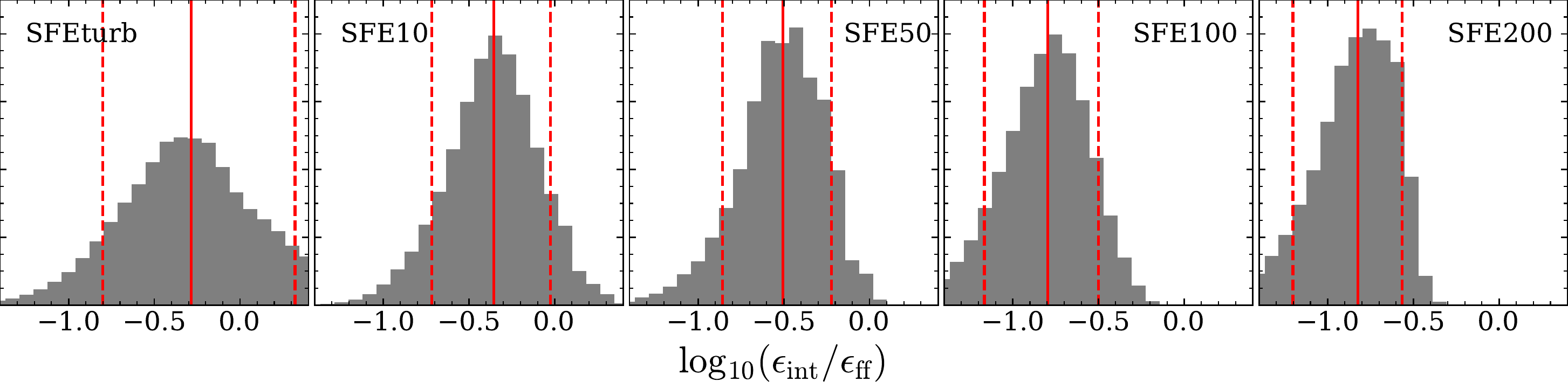}
\caption{\small Distribution of the mass-weighted integrated star formation efficiency $\epsint$ relative to the instantaneous star formation efficiency per freefall time $\epsff$. Red solid and dashed lines show the median and 10th-90th percentile range for each distribution.
}\label{fig:epsint}
  \vspace{0.3cm}
\end{figure*}

The reason for this overestimate is not clear yet. It may be caused by the averaging scheme. The morphology of our simulated galaxies is irregular (see Figure~\ref{fig:gas-proj}), unlike the axisymmetric disks of low-redshift galaxies. Big gaps in the gas density distribution, caused by strong feedback, make the estimate of $\Sigma_{\rm H_2}$ less reliable. Interestingly, the FIRE simulations with spatial resolution comparable to ours show similar overestimate of $\Sigma_{\rm SFR}$, pushing the gas depletion timescale into the range between 100~Myr and 1~Gyr. This depletion timescale, although shorter than the measurements from the local universe, is however not inconsistent with high-$z$ results \citep{sharon_etal13, tacconi_etal13, rawle_etal14, hodge_etal15, sharda_etal18}.

\subsection{Initial bound fraction} \label{sec:initial_bound_result}

Figure~\ref{fig:initial_bound} shows the initial bound fraction predicted for our model clusters. We find a strong positive correlation between $f_i$ and stellar particle mass for all runs with a wide range of model parameters. This trend is not sensitive to either the formation epoch or host galaxy mass, suggesting that $f_i$ is mostly independent of the global galactic environment and reflects only local properties of individual star-forming regions. 

Instead, we find that the initial bound fraction varies strongly with $\epsff$. At $M = 10^5\Msun$, $f_i$ can reach above 50\% for $\epsff \ge 0.5$, but is limited to only $1-10\%$ for runs with $\epsff < 0.5$. This difference in normalization is due to the corresponding scaling of the integral star formation efficiency because of our assumption $f_i \propto \epsint$ (Equation~\ref{eq:initial_bound}).

The dependence of $\epsint$ on stellar particle mass has already been noticed in recent GMC-scale simulations. For example, \citet[][hereafter, G18]{grudic_etal18} ran a series of MHD simulations of isolated turbulent clouds with different initial gas surface density. They found a tight relation between $\epsint$ and $\Sigma_{\rm gas}$ and parameterized it by the relation
\begin{equation} \label{eq:eps_FIRE}
  \epsint = \left( \frac{1}{\epsilon_{\rm max}} + \frac{\Sigma_{\rm crit}}{\Sigma_{\rm gas}} \right)^{-1},
\end{equation}
with best-fit parameters $\epsilon_{\rm max} = 0.77$ and $\Sigma_{\rm crit} = 2800\Msun\, {\rm pc}^{-2}$. The critical surface density $\Sigma_{\rm crit}$ above which the efficiency rises significantly corresponds to the cloud mass $M_{\rm crit}\approx 2.2\times 10^{5}\Msun\, (R/5\,{\rm pc})^2$. Using our definition of the initial bound fraction, the above equation can be rewritten as a relationship between $f_i$ and scaled particle mass $m \equiv M/(\epsilon_{\rm max} M_{\rm crit})$:
\begin{equation} \label{eq:fi_fire}
  f_i = \min{\left( \sqrt{m^2+4m}-m, 1 \right)}.
\end{equation}
We emphasize that the radius $R$ used in G18 is not the same as the radius of our GMC sphere, $\Rgmc$. In their simulations of isolated clouds, both the mass and size of the cloud are fixed by the initial condition, while our star-forming GMC has an open boundary through which gas flows in and out, controlled by gravity and stellar feedback. Our $\Rgmc$ is a lower limit to $R$ since the cluster can accrete more distant gas, e.g. from all 27 neighbor cells. We can estimate a radius of the corresponding accretion region by taking it to be a sphere of the same volume as the 27 gas cells. Since the physical size of our cells at the finest refinement level varies between 3 and 6~pc (see Section~\ref{sec:refinement}), we take the average length to be with 4.5~pc. This gives the effective radius $R \approx 8.4$~pc. We use this value to compare the G18 scaling with ours.

The right panel of Figure~\ref{fig:initial_bound} shows a good agreement between the relation given by Equation~(\ref{eq:fi_fire}) and our runs with $\epsff \ge 1$. The lower efficiency runs cannot reach the relatively high values of $f_i$ predicted by G18. We fit the relation between stellar particle mass and initial bound fraction as $f_i\propto M^{a}$, and find the power-law slope for our clusters in the range $a = 0.43-0.51$. It is consistent with the G18 result $a\approx 0.5$ for clusters in the mass range $M < 10^5\Msun$.

In the SFEturb run, $f_i$ has a nonmonotonic downturn at $M\sim 10^5\Msun$. This downturn is caused by the relatively small number of massive clusters in this run, so that the median value of $f_i$ is dominated by some rare cluster formation events. Indeed, these massive clusters are formed in a small satellite galaxy that contains only one compact GMC with mass $\sim 10^8\Msun$ at $z\sim 4$. The low $\epsff$ in the SFEturb run leads to slow gas consumption, which allows the cloud to persist for a long time. Because of the high density of material in the vicinity of the GMC, young clusters acquire large velocity dispersion and leave the GMC. New clusters then appear in their stead and draw gas from the same cloud. This large cloud mass in the denominator of Equation~(\ref{eq:epsint}) then leads to lower integrated star formation efficiency $\epsint$, which causes the downturn of $f_i$. This effect is less visible for the higher-efficiency runs, where GMCs are consumed and disrupted quicker.

The initial bound fraction in our models depends somewhat on the SNR boosting factor $\fboost$. The run with $\fboost=3$ shows systematically higher $f_i$ than the corresponding $\fboost=5$ run, although only by $\sim20\%$. The dependence of $f_i$ on $\fboost$ can be understood as the balance between gravity and feedback: with lower $\fboost$, a cluster in one GMC of a given mass requires more stellar mass to reach the same amount of SN momentum to fully disperse the gas. Therefore, a cluster in the SFE50-SNR3 run reaches a higher final mass for the same GMC mass, which implies higher values of $\epsint$ and $f_i$.

There are several consequences of introducing the initial bound fraction to determine the initial cluster mass. First, due to the mass dependence of $f_i$, the shape and normalization of the CIMF is different from the stellar particle mass function. Second, the sensitivity of $f_i$ to the choice of $\epsff$ leads to a different integrated cluster formation efficiency and maximum cluster mass for runs with different $\epsff$. These observables can be used to constrain $\epsff$, as we discuss below.

\subsection{Scatter of the integral efficiency of star formation}

Because of the time-dependent accretion of gas onto GMCs and dispersal by stellar feedback, the integral efficiency $\epsint$ of cluster formation deviates from the adopted efficiency per freefall time $\epsff$ in a given run.

Figure~\ref{fig:epsint} shows the particle mass-weighted ($\epsint M$) distributions of $\epsint/\epsff$ in all of our main runs. In the case of the SFEturb run, the instantaneous efficiency $\epsff$ in a given cell depends on the local properties of supersonic turbulence. We calculate the mass-weighted mean efficiency $\langle\epsff\rangle \approx 0.03$ for this run and use it in the leftmost panel, which shows $\epsint/\langle\epsff\rangle$.

There exists a weak trend that the median value of the distribution of $\epsint/\epsff$ decreases with $\epsff$. For the SFEturb, SFE10, SFE50, SFE100, SFE200 runs, the median values of $\epsint/\epsff$ are 0.52, 0.44, 0.31, 0.16, and 0.15, which correspond to the values of $\epsint=0.016$, 0.044, 0.155, 0.16, and 0.3, respectively. Larger $\epsff$ leads to faster initial growth of a given cluster, which produces more intense stellar feedback. Stronger feedback removes more material from the GMC and leads to larger deviation of $\epsint$ from $\epsff$.

Similar to Paper I, all distributions show a large scatter of $\epsint$. The 10th-90th percentile range around the median value is $\approx 0.7\,$dex for all of the runs, except the SFEturb run which has a larger range of $1.1\,$dex. The equivalent $1\,\sigma$ scatter is about 0.25~dex for the fixed $\epsff$ runs, and 0.43~dex for the SFEturb run. The larger scatter in the latter run is caused by the additional variation of instantaneous $\epsff$ over the course of cluster growth because of the variation of local turbulence.

\begin{table}
\centering
\caption{CIMF best-fit parameters}
\begin{tabular}{lcclc}
\tableline
Runs 		& $M_{\rm min}/\Msun$	& $\alpha$	& $M_{\rm cut}/\Msun$\\
\tableline
 & Full-particle & & \\
\tableline\\
SFE10 		& $6\times10^4$	& 3.4	& $1.3\times10^5$	\\
SFE50 		& $6\times10^4$	& 3.3	& $1.6\times10^5$	\\
SFE50-SNR3 	& $4\times10^4$	& 3.7	& $4.5\times10^5$	\\
SFE100	 	& $4\times10^4$	& 3.1	& $4.5\times10^5$	\\
SFE200	 	& $4\times10^4$	& 3.6	& NA$^a$	\\
\tableline
 & Full-cluster & & \\
\tableline\\
SFE10 		& $5\times10^3$	& 2.6	& $3.6\times10^4$	\\
SFE50 		& $2\times10^4$	& 2.5	& $3.1\times10^5$	\\
SFE50-SNR3 	& $1\times10^4$	& 2.9	& $1.3\times10^5$ \\
SFE100	 	& $2\times10^4$	& 2.5	& $2.3\times10^5$	\\
SFE200	 	& $4\times10^4$	& 3.3   & NA$^a$	\\
\tableline
 & $z\approx2.0$-cluster& & \\
\tableline\\
SFE10 		& $3\times10^3$	& 2.6	& $2.4\times10^4$	\\
SFE50 		& $8\times10^3$	& 2.2	& $4.0\times10^4$	\\
SFE50-SNR3 	& $1\times10^4$	& 2.0	& $1.3\times10^4$	\\
SFE100	 	& $1\times10^4$	& 1.7	& $3.5\times10^4$	\\
SFE200	 	& $6\times10^3$	& 1.2	& $2.5\times10^4$	\\
\tableline
 & $z\approx5.3$-cluster & & \\
\tableline\\
SFE10 		& $2.5\times10^3$ & 1.1	& $1.6\times10^4$	\\
SFE50 		& $6\times10^3$	& 1.3	& $1.3\times10^5$	\\
SFE50-SNR3 	& $6\times10^3$	& 1.8	& $1.7\times10^5$	\\
SFE100	 	& $5\times10^3$	& 1.4   & $2.5\times10^5$	\\
SFE200	 	& $3\times10^4$	& 1.5 	& $8.5\times10^5$	\\
\tableline
\end{tabular}
 \label{tab:CIMF}\\
 \vspace{0.1cm}
Note: $a.$ The cutoff mass cannot be obtained, since the CIMF of all clusters in the SFE200 run is consistent with a pure power law.
\end{table}

\subsection{CIMF} \label{sec:result-CIMF}

The CIMF is one of the key properties of young star clusters. Here we examine the CIMF of model clusters in the main halo, summed over the central galaxy and satellite galaxies. In Paper I we showed that the Schechter function provides a good description of the shape of the cluster mass function. Here we also fit all CIMFs with a Schechter function using the maximum-likelihood method. Because the shape of the CIMF for small clusters now deviates more from a single power law, we restrict the fit only to clusters more massive than a certain minimum mass, above which the CIMF is best described by the Schechter form. The best-fit slopes, cutoff masses, and choice of minimum mass at different epochs for all runs are listed in Table~\ref{tab:CIMF}. The typical $1\sigma$ uncertainty for the Schechter fit is around $0.1-0.15$ for $\alpha$ and around $0.06-0.13$ dex for $M_{\rm cut}$.

\begin{figure}[t]
\includegraphics[width=1.0\hsize]{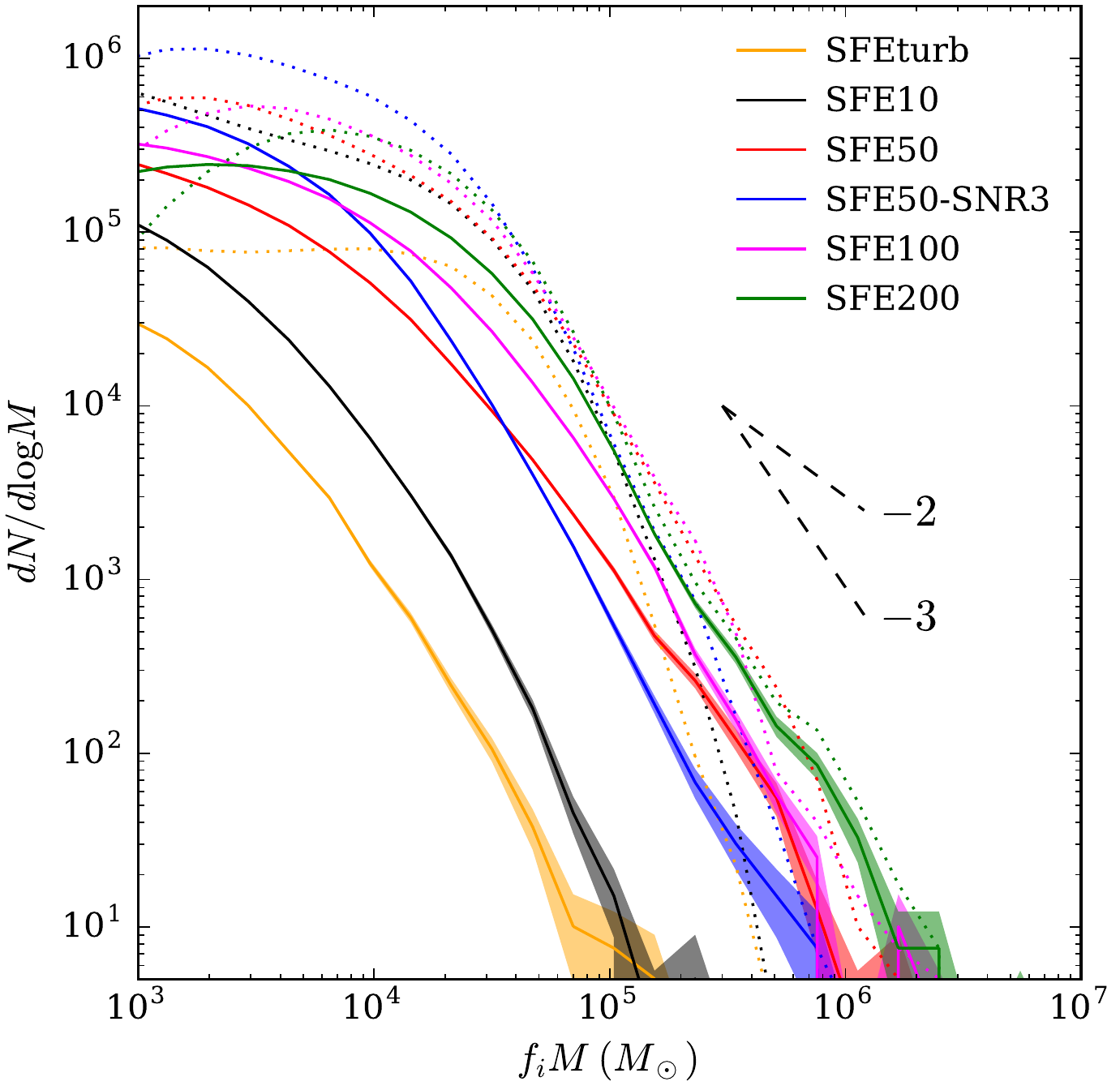}
\caption{\small The CIMF of all clusters ($f_i\,M$; solid lines) within the main galaxy from the last available snapshot of each run. Shaded areas show the binomial counting errors in mass bins of 0.16~dex. In contrast, dotted lines show the distribution of stellar particle mass ($M$) without considering the initial bound fraction. The power-law distributions with slope $\alpha = -2$ and $-3$ are overplotted as dashed lines for reference.
}\label{fig:CMF_full}
\end{figure}

In Figure~\ref{fig:CMF_full}, we show the combined CIMF of all clusters formed at all times up to the last available snapshot of a given run. Here we distinguish between the stellar particle mass ($M$) and the cluster mass that takes into account the initial bound fraction ($f_i\, M$). Because $f_i$ itself depends positively on particle mass, the power-law slope of the CIMF is in general shallower than that of the particle mass function. Among different runs, we find that the high-mass end of the CIMF is also strongly affected by the initial bound fraction. Simulations with lower $\epsff$ tend to have lower $f_i$ for a given cluster mass, thus modifying the mass function more strongly at all masses. We notice that the CIMFs here extend only to $\sim 10^6\Msun$ for $\epsff\geq 0.5$ runs, rather than $\sim 10^7\Msun$ as in Paper I. This is mainly due to the lower SFRs caused by stronger feedback implementation in the current simulations. The maximum mass is reduced to $\sim 10^5\Msun$ for the SFEturb and SFE10 runs because of even smaller values of $f_i$.

\begin{figure}[h]
\includegraphics[width=1.0\hsize]{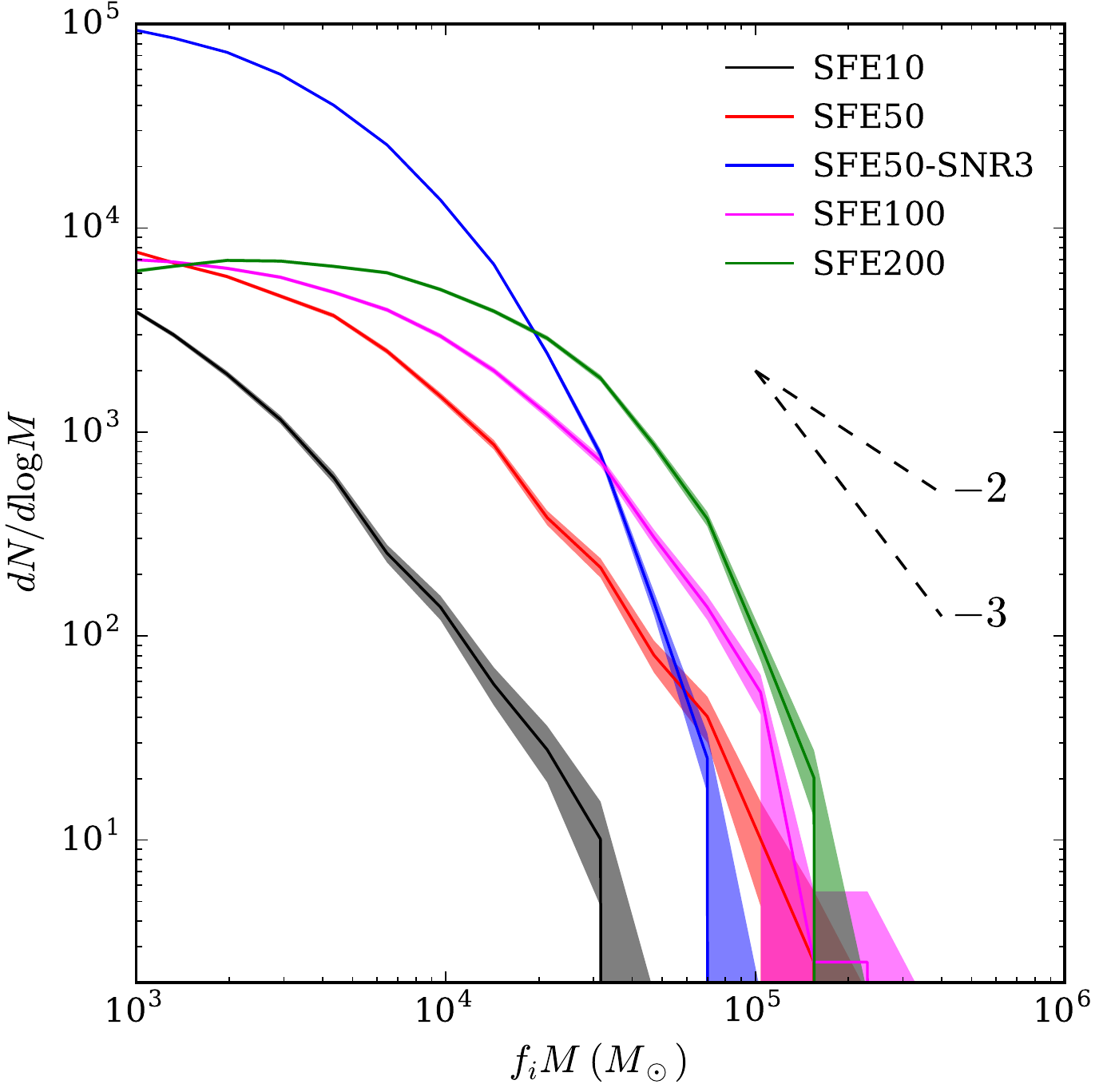}
\caption{\small Same as Figure~\ref{fig:CMF_full}, but only for clusters younger than 100~Myr within the main galaxy at a quiescent epoch around $z \approx 2$.
}\label{fig:CMF_33}
\vspace{0.1cm}
\end{figure}

\begin{figure}[t]
\includegraphics[width=1.0\hsize]{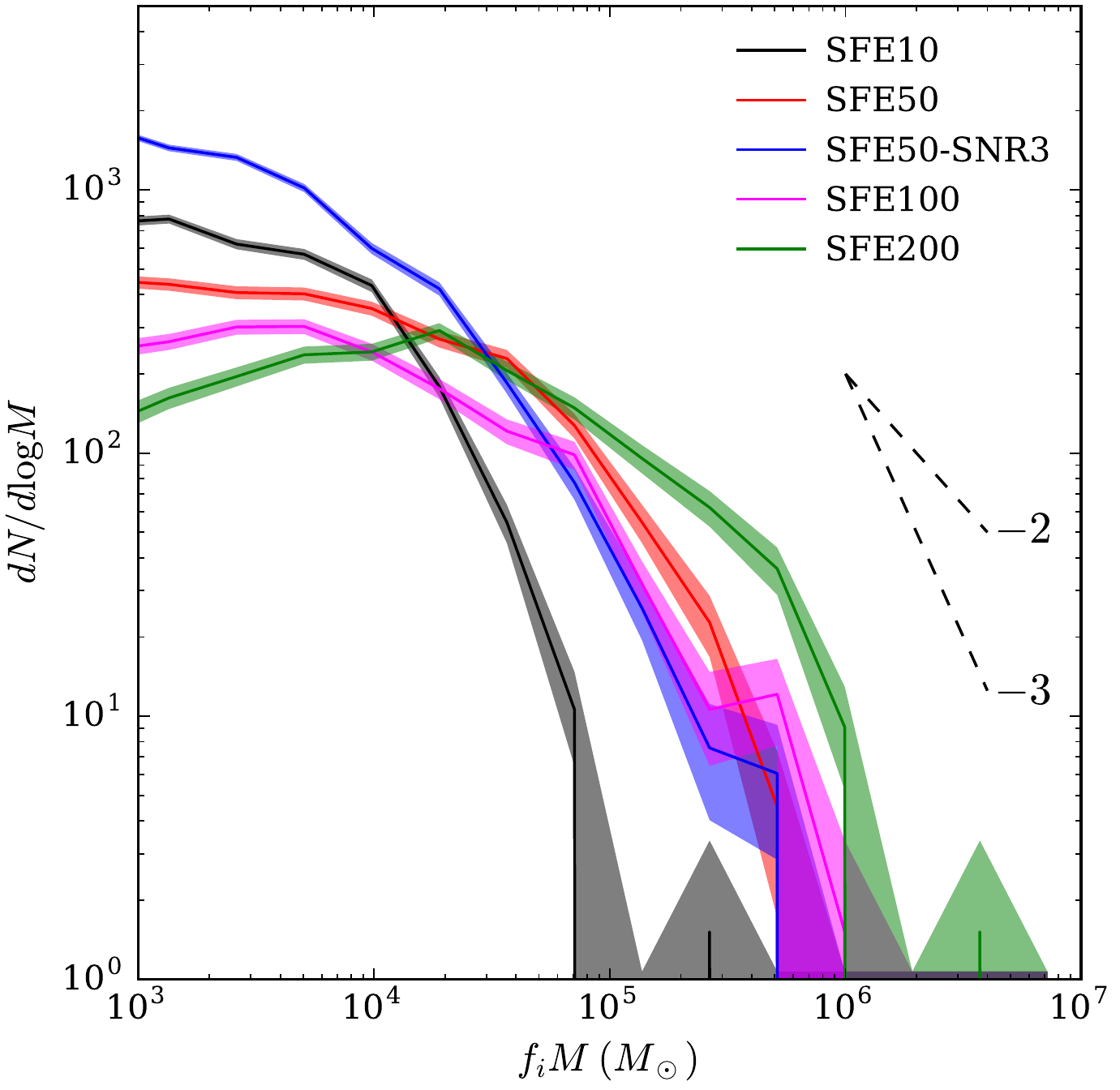}
\caption{\small Same as Figure~\ref{fig:CMF_33}, but during a major merger epoch at $z \approx 5.3$.
}\label{fig:CMF_16}
\end{figure}

To compare more directly with observations of young clusters formed in the same star formation episode, in Figures~\ref{fig:CMF_33} and \ref{fig:CMF_16} we show the CIMF of clusters younger than 100 Myr at different epochs, $z\approx2$ and 5.3, respectively. At $z\approx2$, the main galaxy has not experienced any major mergers for more than 500~Myr. The CIMF of young clusters reflects the properties of the ISM with quiescent star formation. We find a wide range of the power-law slopes for different runs, from $\alpha \approx 2.6$ for the SFE10 run to $\alpha \approx 1.2$ for the SFE200 run. There exists a systematic trend that higher $\epsff$ leads to smaller $\alpha$. A shallower CIMF slope means that, for a given galactic SFR, clusters with higher mass are more likely to be created. That is the reason why the runs with higher $\epsff$ tend to have higher-mass tails in the CIMF.

Figure~\ref{fig:CMF_16} shows the CIMF of young clusters around $z\approx 5.3$, when the main galaxy experiences a major merger. The overall normalization of the mass function is lower than that at $z\approx2$, because the galaxy at $z\approx 5.3$ is much smaller and contains less gas. However, the maximum cluster mass at this epoch is roughly one order of magnitude higher than that at $z\approx2$. This is because the power-law slopes of the CIMF, $\alpha \approx 1.1-1.8$, are much shallower than those in the non-merger case. The cutoff masses of these CIMFs are all above $10^5\Msun$, which is several times larger than those at $z\approx 2$. We already saw such an enhancement of the formation of massive clusters in gas-rich galaxy mergers in the previous runs presented in Paper~I. This effect is even more pronounced in the new runs with stronger feedback.

\subsection{Fraction of clustered star formation} \label{sec:result-Gamma}

In Paper~I, we concluded that cluster formation is an environmentally dependent process where the high-mass end of the CIMF varies strongly with the star formation activity of the host galaxy. Both recent observations \citep{adamo_etal15, johnson_etal16, johnson_etal17} and theoretical work \citep{kruijssen15} suggest that SFR surface density ($\SigmaSFR$), rather than SFR itself, better represents the intensity of star formation and physical conditions of the galactic ISM. Therefore, here we explore the effects of $\SigmaSFR$ on such cluster properties as the fraction of clustered star formation, $\Gamma$, and maximum cluster mass, $M_{\rm max}$.

Analogous to constructing the KSR in Section~\ref{sec:KS}, we split the disk of the main galaxy into the $1\times1$~kpc square grid. For each square area, we calculate the surface density of the SFR averaged over 20~Myr. The fraction of clustered star formation is defined as the ratio between the mass in bound clusters and the total stellar mass formed within the same time interval, 20~Myr:
\begin{equation}
  \Gamma \equiv \frac{\sum f_i M }{\sum M}.
\end{equation}
We also test a larger averaging timescale from 20 to 50~Myr and find that the results are not sensitive to the choice of the timescale.
 
This $\Gamma$ is different from the quantity we used in Paper I in two ways. First, we estimated $\Gamma$ in Paper~I by splitting the galactic disk into concentric circular bins. However, due to the stronger feedback implemented in this paper, no well-defined gas disk is present in the main galaxy. Instead, the disks are clumpy and asymmetric, often showing features of strong outflows. Therefore, it is not possible to find a definitive center and cylindrical-symmetric axis to construct circular bins. Second, in Paper~I, we defined $\Gamma$ simply as the fraction of clusters more massive than $10^4\Msun$, not all of which are necessarily gravitationally bound. In this paper, with the introduction of the initial bound fraction, we have a more physical way of estimating the clustered fraction as it is defined in observations.

\begin{figure}[t]
\includegraphics[width=1\hsize]{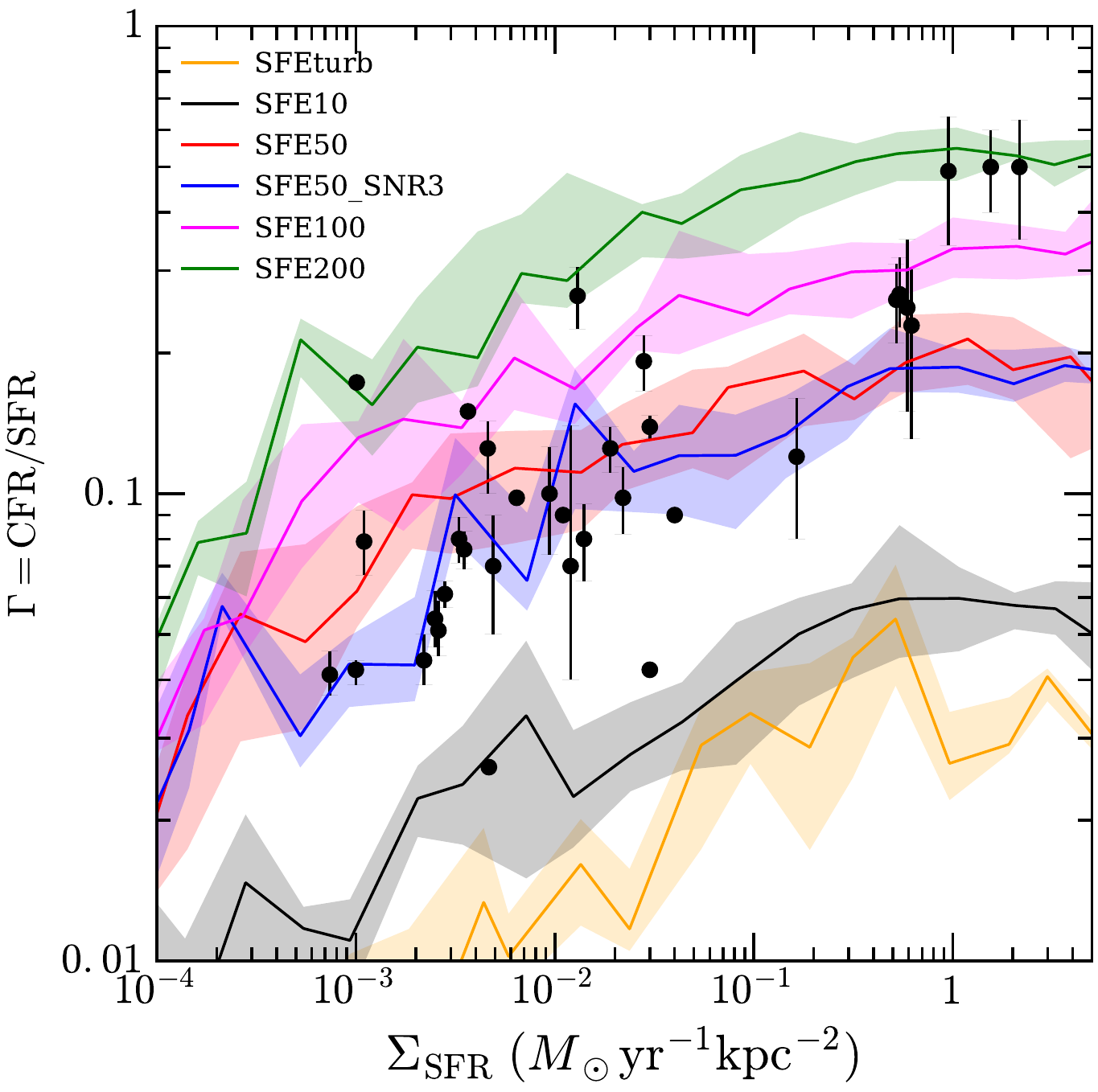}
\caption{\small
Fraction of clustered star formation as a function of SFR surface density. Here $\SigmaSFR$ is estimated on a spatial scale of 1~kpc for stars younger than 20 Myr. Solid lines and shaded areas show the median and 25-75\% interquartile range of the distribution of $\Gamma$ for a given $\SigmaSFR$ bin. The observed values (symbols with error bars) are from a compilation of both galaxy-wide and spatially resolved measurements of cluster samples in nearby galaxies \citep{goddard_etal10, silva-villa_larsen11, adamo_etal15, johnson_etal16}.
}\label{fig:Gamma}
  \vspace{1mm}
\end{figure}

Figure~\ref{fig:Gamma} shows a positive correlation between $\Gamma$ and $\SigmaSFR$ for all the runs. We find that $\Gamma$ changes by about one order of magnitude over the range $\SigmaSFR = 10^{-3}-1 \Msun\, \mathrm{yr}^{-1}\, \mathrm{kpc}^{-2}$. At $\SigmaSFR > 1 \Msun\, \mathrm{yr}^{-1}$, $\Gamma$ saturates at the maximum value set by the initial bound fraction of the representative massive clusters. A similar saturation of $\Gamma$ at high $\SigmaSFR$ is also found in the analytical model of \citet{kruijssen_etal12}, however, the physical origin of this saturation is different. In \citet{kruijssen_etal12}, the saturation is caused by the ``cruel cradle effect,'' in which young clusters formed within less-dense regions are destroyed by the tidal interaction with other star forming regions during their embedded phase, while in our model, the saturation is caused by the mass-dependent initial bound fraction.

It is clear that $\Gamma$ also depends strongly on $\epsff$. Run SFE200, with $\epsff=2.0$, shows a very high $\Gamma$ up to 60\%, while runs SFE10 and SFEturb do not have any star-forming regions with $\Gamma > 10\%$. This sensitivity makes $\Gamma$ an excellent observable to constrain the choice of $\epsff$.

\subsection{Maximum cluster mass}\label{sec:result-mmax}

\begin{figure}[t]
\includegraphics[width=1.0\hsize]{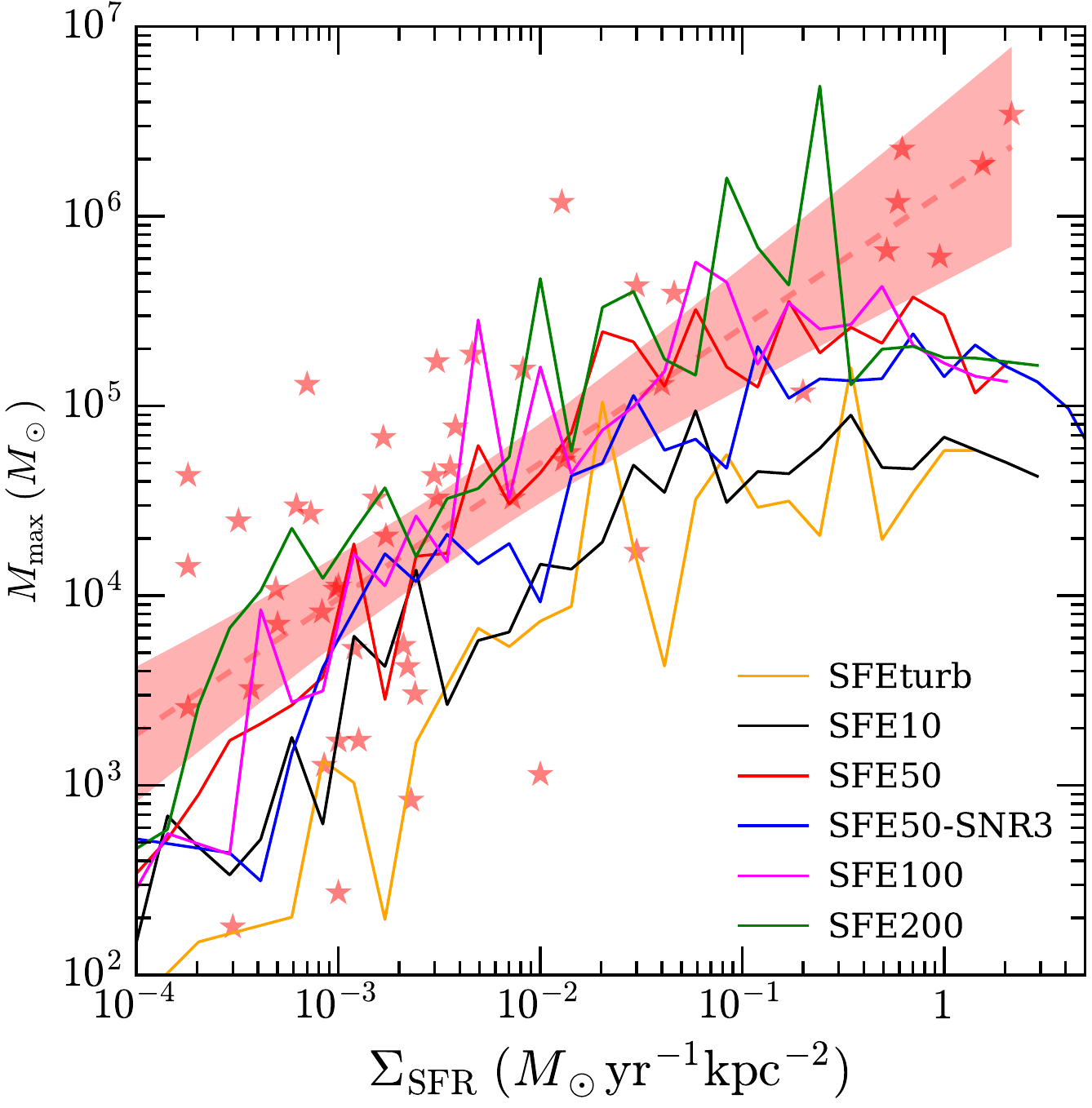}
\caption{\small 
Maximum bound cluster mass vs. SFR surface density. Here $\SigmaSFR$ is calculated on the 1~kpc scale for clusters younger than 100~Myr across the disk of the main galaxy from z=10 to the last available snapshots for different runs. The compilation of observed maximum cluster masses in different galaxies from Appendix B in \citet{adamo_etal15} is shown with red stars. The best linear fit to the data is overplotted as a red dashed line, along with its 1$\,\sigma$ confidence interval (red shaded area).
}\label{fig:sfh-mmax}
\end{figure}

Figure~\ref{fig:sfh-mmax} shows the relationship between $\SigmaSFR$ and the maximum mass of clusters formed within a 100~Myr interval. We find a clear positive correlation going roughly as
$$ M_{\rm max} \propto \SigmaSFR^{2/3}. $$
The power-law slope is similar to the best-fit value, $\approx 0.7$, for the observations of maximum cluster mass described in Section~\ref{sec:Mmax-obs}. The similarity of the slope among all runs reveals a robust result that the high-mass end of the CIMF depends uniquely on the intensity of star formation activity. The normalization of this relation also scales monotonically with $\epsff$, because of the $\epsff$ dependence of the initial bound fraction.

Note that the runs with $\epsff\le 0.1$ cannot produce clusters more massive than $10^5\Msun$ at all considered epochs, $z>1.5$. This indicates that no clusters in these runs would survive dynamical disruption to become globular clusters at the present time. This is another evidence against such low values of $\epsff$.

\begin{figure*}[t]
\centerline{\includegraphics[width=0.7\hsize]{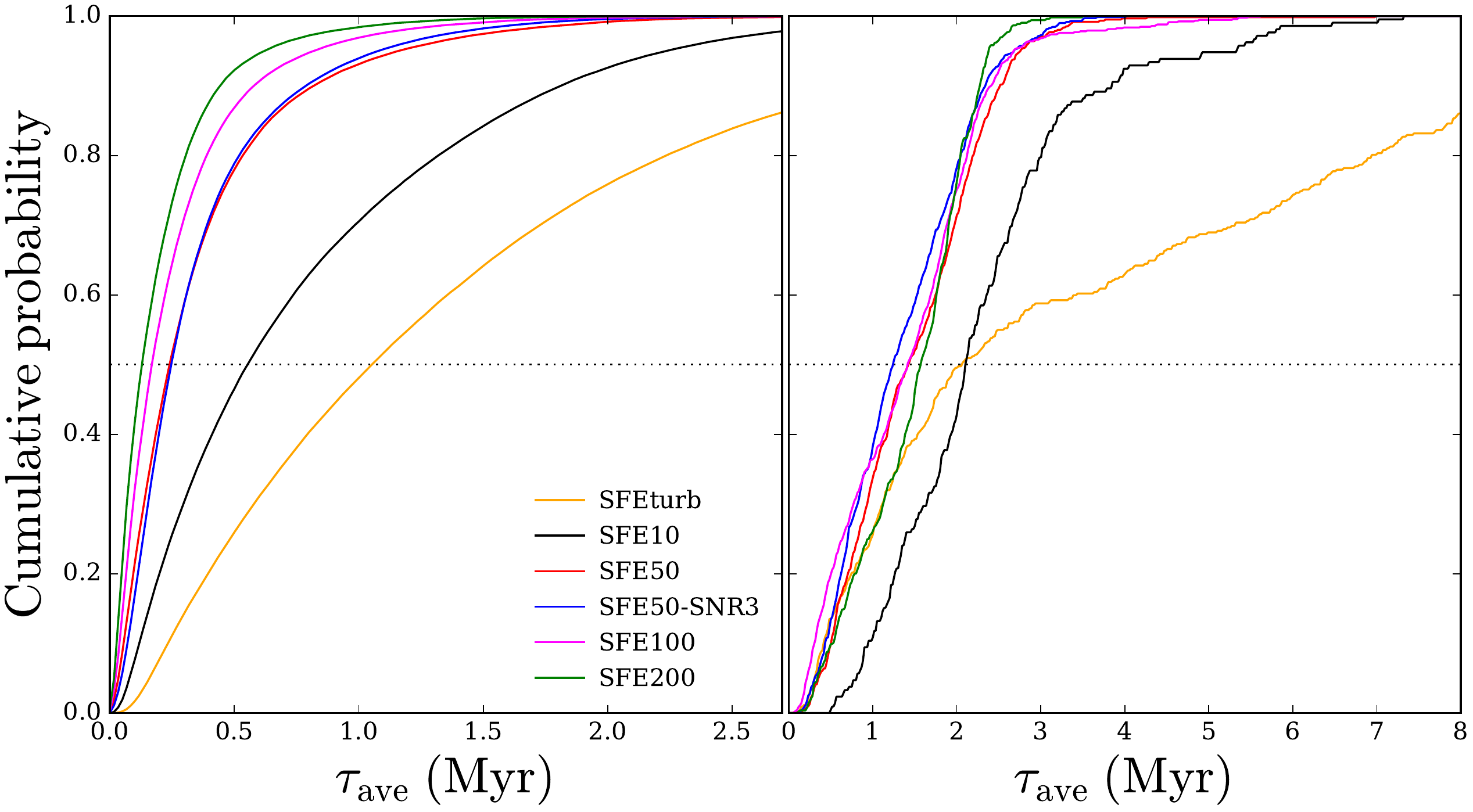}}
\centerline{\includegraphics[width=0.7\hsize]{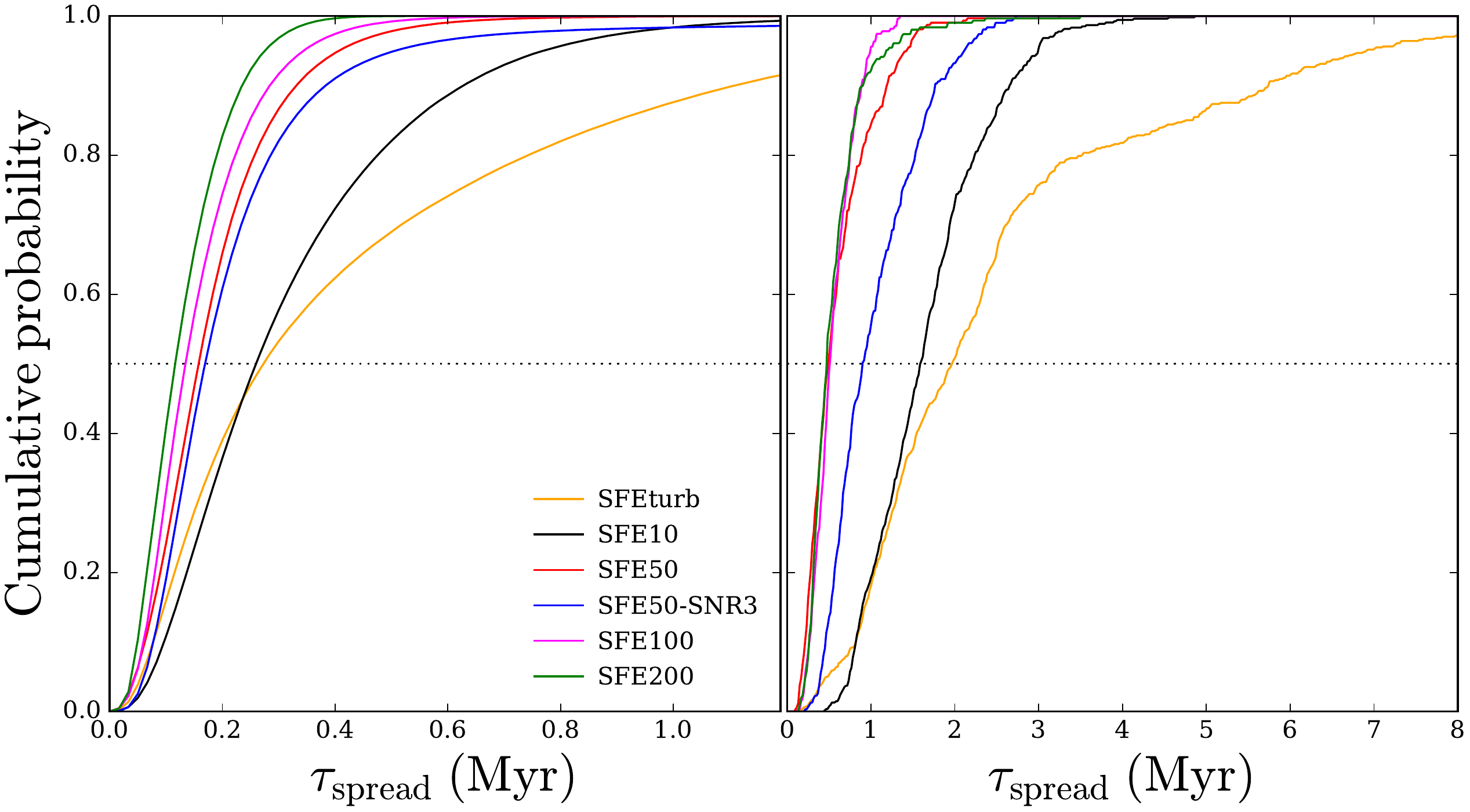}}
\caption{\small Cumulative distribution of mass-weighted cluster formation timescale $\tauave$ (\textit{upper panels}) and newly defined age spread $\tauspread$ (\textit{lower panels}) for clusters with masses smaller (\textit{left panels}) and larger (\textit{right panels}) than $10^5\Msun$.
}\label{fig:cft}
  \vspace{1mm}
\end{figure*}

\subsection{Cluster formation timescale} \label{sec:result-tau}

The upper panels of Figure~\ref{fig:cft} show the cumulative distributions of cluster formation timescales $\tauave$. We find a clear trend that the higher the $\epsff$, the shorter the timescale, especially for low-mass clusters ($M< 10^5\Msun$). This means that the growth history of these clusters is dominated by the gas accretion, which is in turn controlled by $\epsff$. This result, obtained after many algorithmic updates described in Section~\ref{sec:updates}, is still consistent with that in Paper~I.

\begin{table}
\centering
\caption{Cluster Formation Timescales in Myr}
\begin{tabular}{lcc}
\tableline\\[-2mm]
Runs 		& $M<10^5\Msun$ & $M>10^5\Msun$ \\
            & $10\%-50\%-90\%$ & $10\%-50\%-90\%$ \\
 \tableline
 & $\tauave$ &\\
  \tableline\\[-2mm]
SFEturb		& $0.25-1.07-3.13$ & $0.43-2.07-8.62$ \\
SFE10		& $0.14-0.57-1.83$ & $0.96-2.12-3.85$ \\
SFE50		& $0.07-0.26-0.84$ & $0.52-1.44-2.53$ \\
SFE50-SFR3	& $0.09-0.26-0.80$ & $0.45-1.26-2.35$ \\
SFE100		& $0.06-0.19-0.60$ & $0.32-1.43-2.42$ \\
SFE200		& $0.04-0.15-0.46$ & $0.55-1.59-2.29$ \\
 \tableline
 & $\tauspread$ &\\
  \tableline\\[-2mm]
SFEturb		& $0.09-0.29-1.13$ & $0.84-1.99-5.76$ \\
SFE10		& $0.11-0.28-0.64$ & $0.84-1.61-2.67$ \\
SFE50		& $0.08-0.17-0.35$ & $0.23-0.51-1.20$ \\
SFE50-SFR3	& $0.09-0.19-0.40$ & $0.48-0.91-1.78$ \\
SFE100		& $0.08-0.15-0.30$ & $0.28-0.53-0.90$ \\
SFE200		& $0.07-0.13-0.25$ & $0.28-0.49-0.91$ \\
\tableline\\[-2mm]
\end{tabular}
 \label{tab:tau}
\end{table}

More quantitatively, we split the whole cluster sample into low- ($<10^5\Msun$) and high- ($>10^5\Msun$) mass clusters and calculate the 10\%, 50\%, and 90\% percentiles of the distribution of $\tauave$. The results for both low- and high-mass clusters are listed in Table~\ref{tab:tau}.

In general, the timescales for low-mass clusters are very short. The median values of $\tauave$ are all below $\sim 1$~Myr. To better understand the changes across different runs, instead of just comparing the median values, we quantitatively evaluate the differences of the probability density distributions of $\tauave$. We introduce a "shift factor" $f_{\rm best}$ that would make cumulative distributions of samples $s_1$ and $s_2$ most similar to each other. Specifically, we iteratively solve for the factor $f_{\rm best}$ that maximizes the $p$-value of the Kolmogorov-Smirnov test between $s_1$ and $f_{\rm best}\, s_2$. We find that the $\tauave$ distributions in the SFE200, SFE100, and SFE50 runs are shifted by 0.27, 0.34, and 0.45, respectively, relative to the SFE10 run. This suggests a strong anticorrelation between $\epsff$ and $\tauave$, with a scaling relation that can be best described as $\tauave \propto \epsff^{-0.45}$. This is interesting because, although the instantaneous gas accretion rate of clusters is linearly correlated with $\epsff$ in Equation~(\ref{eq:sfr}), the correlation between $\epsff$ and $\tauave$ is nonlinear.

On the other hand, the comparison of the distributions of $\tauave$ for the SFE50 and SFE50-SNR3 runs shows that the boosting factor of SN feedback $\fboost$ has a negligible effect. This is expected, since $\fboost$ only controls the intensity of the momentum feedback from SNe that explode after 3~Myr, which is longer than the typical formation timescale. Feedback from SNe marks the very end of the formation of low-mass clusters.

For massive clusters ($M>10^5\Msun$), however, the distributions of the cluster formation timescale are very similar for runs with $\epsff \geq 0.5$. They all have a median timescale around 1.5~Myr and truncate at $\tauave \approx 3$~Myr, the time when the first SNe explode. This suggests that the growth history of massive clusters is determined by both gas accretion and stellar feedback. However, an immediate termination of gas accretion by SNe only happens with high $\epsff$. Slower star formation cannot accumulate enough massive stars and SNe to disperse the natal GMCs within the first 3~Myr. This can be seen from the long tail of $\tauave$ in SFE10 and SFEturb runs. Although the median value is only $\approx 2-3$~Myr, there are many clusters with $\tauave > 3$~Myr in the SFE10 run and $\tauave > 5$~Myr in the SFEturb run. Such long timescales are inconsistent with the observed age spread of young star clusters in nearby star-forming regions, as discussed in Section~\ref{sec:cft-obs}.

The alternative definition of the formation timescale, $\tauspread$, shows even smaller values than $\tauave$. The median age spread ranges from 0.1 to 0.3~Myr for low-mass clusters and from 0.5 to 2.0~Myr for high-mass ones (see Table~\ref{tab:tau}). However, the cumulative distribution of $\tauspread$ for the SFEturb run still has a long tail toward a large age spread, which is again inconsistent with the observations. In general, $\tauspread$ correlates with $\epsff$, but less strongly than $\tauave$.

\section{Discussion} \label{sec:discussion}

We can now discuss the constraints on $\epsff$ and $\fboost$ resulting from the tests described above.

\subsection{Global properties: SFH and KSR}

In Section~\ref{sec:sfh} we found that the SFH of the main galaxy in runs with $\fboost=5$ is consistent with the abundance matching results. This SFH is very sensitive to the choice of $\fboost$, which gives a strong constraint on its value: $3 < \fboost < 10$. 

On the other hand, the global SFH depends little on the value of $\epsff$ over the whole two orders of magnitude range. The inefficiency of star formation on large scales is not set by hand using small values of instantaneous $\epsff$. Rather, as proposed recently by \citet{semenov_etal17}, the inefficiency may come from multiple star formation cycles on small scales, where stellar feedback processes quickly destroy individual star-forming regions. As long as the feedback is strong enough to trigger these star formation--feedback cycles, the global properties of galaxies and the inefficient star formation activity can be reproduced, and the outcome is not sensitive to the detailed implementation. This phenomenon is also found in high-resolution cosmological simulations by \citet{hopkins_etal14}.

In contrast to the good match of the SFH, the normalization of the KSR in all of our runs is overestimated by a factor of 2-20. In the analysis, we used the 20 Myr averaging timescale to calculate $\SigmaSFR$. We chose this short timescale because we found that the structure of the ISM in our simulations is vulnerable to the feedback from young clusters, so that even the kpc-scale gas distribution changes very rapidly. Also, since we used the high density threshold for cluster formation, $n_{\rm th} = 10^3\,{\rm cm}^{-3}$, the star-forming regions are concentrated in the densest parts of the galaxy. Therefore, the $1\times1$ kpc grid used to derive the KS relation sometimes contains only a couple of star-forming complexes. This situation is very different from the observations of low-redshift galaxies, where comparable areas contain many small star-forming regions, and both $\Sigma_{\rm H_2}$ and $\SigmaSFR$ are averaged over different phases of star formation.

Analogous to the instantaneous SFR in Equation~(\ref{eq:sfr}), higher $\epsff$ runs tend to have higher $\SigmaSFR$ for a given $\Sigma_{\rm H_2}$. However, the slope of the molecular KSR is nearly linear and consistent with the observed linear relation, despite the nonlinear scaling $\dot{\rho}_*\propto \rho_{\rm gas}^{3/2}$ used in Equation~(\ref{eq:sfr}).

\subsection{Slope of the star cluster mass function}

In Section~\ref{sec:result-CIMF}, we showed that the power-law slope of the CIMF is sensitive to the choice of $\epsff$. Observations of the CIMF in nearby galaxies reveal a Schechter-like function with a power-law slope close to $-2$ \citep{portegies_zwart_etal10}. The SFE10 run shows the slope as steep as $\approx 2.5$, while the SFE200 run has a very shallow slope smaller than 1.5. Therefore, under the current simulation setup, we find that the range $\epsff = 0.5-1.0$ matches the observations best.

The normalization of the CIMF also depends strongly on $\epsff$.  Because of the positive correlation between $\epsff$ and the initial bound fraction, runs with lower $\epsff$ create clusters that have lower $f_i$ for a given particle mass. Therefore, applying the initial bound fraction shifts the particle mass function to smaller masses and leads to the apparent lower normalization of the CIMF for lower $\epsff$ runs.

This conclusion relies on the adopted linear model for the initial bound fraction, Equation~(\ref{eq:initial_bound}), which is an extrapolation of a few simulation results available in the literature \citep{kruijssen12}. If the true relation is not linear or has a large scatter, as indicated recently by \citet{goodwin09, dale_etal14, gavagnin_etal17}, the derived CIMF would be affected. The overall galaxy dynamics is unaffected by the model for $f_i$ because we assume that the bound and unbound components of a given cluster move together.

\subsection{Effects of major mergers}

In Section~\ref{sec:result-CIMF}, we find that there is a significant difference in CIMFs produced during major mergers and during quiescent periods. In a merger, the slope of the CIMF becomes shallower, while the truncation mass rises much higher. This promotes the formation of most massive clusters, $M\ga 3\times 10^5\Msun$. It is similar to the result we found in Paper I, despite the many updates to the algorithm in this paper.

The existence of more than 100 globular clusters in the Milky Way at present provides an additional constraint on the creation of most massive clusters at high redshift. As can be seen in Figures~\ref{fig:CMF_full} and \ref{fig:sfh-mmax}, clusters with masses larger than $\sim 3\times 10^5\Msun$, which are potential globular cluster progenitors, can only be created in the runs with $\epsff\ge 0.5$.

\subsection{Fraction of clustered star formation}

In Paper~I, we presented a positive correlation between $\SigmaSFR$ and the fraction of young clusters with mass above $10^4\Msun$. Roughly speaking, this fraction is a reasonable proxy for the fraction of clustered star formation that we estimated in Section~\ref{sec:initial_bound_result}. This is because massive clusters tend to have higher $f_i$, and therefore, most of the bound cluster mass is contributed by the most massive clusters. In Paper~I the shape of the particle mass function was not affected strongly by the value of $\epsff$, and similarly, the fraction of massive clusters was not sensitive to the choice of $\epsff$. In the new runs, however, the strong correlation between $\epsff$ and the initial bound fraction $f_i$ leads to a positive correlation between $\epsff$ and the integrated cluster fraction, $\Gamma$. Thus, $\Gamma$ as a function of $\SigmaSFR$ provides us with a new diagnostic to constrain $\epsff$.

The positive correlation between $\SigmaSFR$ and $\Gamma$ found in Section~\ref{sec:result-Gamma} is caused by two effects. First, in high-$\SigmaSFR$ areas, the CIMF is likely to extend to higher masses, as demonstrated in Figure~\ref{fig:sfh-mmax}. Second, high-mass clusters typically have larger initial bound fractions, which leads to higher $\Gamma$. Therefore, this correlation comes about from a combination of the variation of CIMF in different environments and the mass-dependent initial bound fraction.

We find that the simulations with $\epsff=0.5-1.0$ reproduce the observed values of $\Gamma$ over a wide range of $\SigmaSFR = 10^{-3}-1 \Msun\, \mathrm{yr}^{-1}\, \mathrm{kpc}^{-2}$. However, there are three data points from \citet{adamo_etal15} that show very high $\Gamma \sim 0.5$ at $\SigmaSFR \sim 1\Msun\, \mathrm{yr}^{-1}\, \mathrm{kpc}^{-2}$, which cannot be reached by either the SFE50 or SFE100 run. It is important to keep in mind that these three data points are from galaxy-integrated rather than spatially resolved measurements, which we did in this paper. Note also that the data compilation shown in Figure~\ref{fig:Gamma} combines measurements of different SFR tracers, different cluster identification criteria, and different averaging spatial and temporal scales. Future observations with consistent methodology and higher spatial resolution are required to place better constrains on the star formation models.

\subsection{Maximum cluster mass}

In Section~\ref{sec:result-mmax}, we find a positive correlation between the SFR surface density and maximum mass of young clusters. The normalization of this relation scales with $\epsff$. We find that runs with $\epsff \ge 0.5$ are consistent with the observations of clusters in nearby galaxies over a large range of $\SigmaSFR = 10^{-4}-2 \Msun\, \mathrm{yr}^{-1}\, \mathrm{kpc}^{-2}$. Due to the small size of observational samples, it is hard to constrain $\epsff$ into a narrower range, but at least $\epsff \le 0.1$ is ruled out.

\subsection{Mass accretion history inferred from different definitions of cluster formation timescale}

In Section~\ref{sec:updates-tau}, we discussed how the two definitions of the cluster formation timescale probe different mass accretion behavior of the clusters.

First, we check whether the mass accretion history can be described by a simple power law, $\dot{M}\propto t^{\alpha}$, as suggested by \citet{murray_chang15}. Such power-law mass accretion gives $\tauave/\taumax = (\alpha+1)/(\alpha+2)$ and \mbox{$\tauspread/\taumax = (2\alpha+1)/(\alpha+1)^2$}. Therefore, if the accretion history of individual clusters can indeed be described by a power law, the indexes $\alpha$ derived from the ratios $\tauave/\taumax$ and $\tauspread/\taumax$ should be consistent with each other. However, we find that majority of the clusters do not show consistent $\alpha$ from these two definitions, suggesting that most clusters experience accretion histories that cannot be described by a simple power law.

Then we explore detailed the output of the mass accretion rate at each local timestep ($\sim 100\,$yr) for a fraction of massive clusters above $10^5\Msun$. We find a general pattern that the accretion rate is relatively low at the beginning of cluster formation, when the gas density is near the threshold $n_{\rm crit}$. As the parent GMC collapses to higher density, the central gas cell where the cluster resides is refined to one or two higher levels, which increases the mass growth rate. We find that this collapse phase typically lasts for $\sim2/3$ of the total cluster formation duration. Such superlinear mass growth is qualitatively consistent with the theoretical prediction of the collapse of turbulence clouds \citep{murray_chang15, murray_etal17}. As the mass accretion reaches its peak by either exhausting the star-forming gas or removing it by stellar feedback, the density quickly declines, and, once it is below the threshold $n_{\rm crit}=10^3{\rm cm}^{-3}$, the cluster growth is shut down completely. This pattern clearly cannot be fitted by a single power law of time, and it is better described by a peaked distribution. In this case, $\tauave$ and $\tauspread$ reflect distinct characteristics of the accretion of individual clusters. While $\tauave$ indicates the timescale for a given cluster to reach the peak of its mass accretion, $\tauspread$ quantifies the width of that peak.

\subsection{Cluster formation timescale}

Similar to the results of Paper~I, in the new runs, the cluster formation timescale depends strongly on $\epsff$. This robust relationship could be used to constrain $\epsff$. In practice, however, the clusters in most runs have formation timescales that are shorter than 3~Myr, so that they all are within the range of the observed age spread. One exception is the SFEturb run, in which $\sim 30\%$ of massive clusters have $\tauave > 6\,$Myr, which is inconsistent with observations.

\subsection{Combination of all constraints} \label{sec:all}

Here we summarize the constraints for the star formation and feedback parameters, $\epsff$ and $\fboost$, resulting from the different observables discussed above:
\begin{itemize}
\item global SFH: $3 < \fboost < 10$.
\item slope of CIMF: $\epsff=0.5-1.0$.
\item fraction of clustered star formation: $\epsff=0.5-2.0$.
\item maximum cluster mass: $\epsff\ge 0.1$.
\item cluster formation timescale: $\epsff\ge 0.1$.
\end{itemize}
Based on the joint constraints, we find that $\fboost \approx 5$ and $\epsff=0.5-1.0$ are favored in the current framework of star formation and stellar feedback subgrid models and the current spatial and mass resolution.

We emphasize again that the $\epsff$ used in this paper is different from that in other cosmological simulations: (1) it is only applied within a star-forming sphere of the fixed physical volume, rather than a time-variable cell volume; and (2) the mass growth rate of a given cluster varies significantly during the active accretion period due to changes of the local gas density, which leads to broad scatter of the integrated efficiency $\epsint$ at a fixed $\epsff$.

\section{Summary} \label{sec:summary}

We have described an improved implementation of star cluster formation in the cosmological code ART. We eliminated the accretion gaps during the period of active cluster growth, which significantly reduced the long formation timescales found in Paper~I. We adopted a new SNR feedback model that brought the global SFH of the main galaxy into agreement with the abundance matching result. We also introduced a new prescription for the initial bound fraction of individual clusters. 

We performed a series of cosmological simulations of a Milky Way-sized galaxy with different star formation and feedback parameters and used various properties of the model clusters to constrain these parameters. The conclusions from this new suite of simulations are summarized here.
\begin{itemize}
\item Global galactic properties, such as the morphology and SFH, are strongly affected by the strength of momentum feedback but are almost insensitive to the star formation efficiency $\epsff$ when the feedback is sufficiently strong.

\item To match the SFH expected from abundance matching for a Milky Way-sized galaxy, the momentum boosting factor of SNR feedback is constrained to be in the range $3 < \fboost < 10$.

\item The initial bound fraction $f_i$ increases with stellar particle mass at all formation epochs and host galaxy masses. The value of $f_i$ also positively correlates with $\epsff$.

\item The CIMF of model clusters can be described by a Schechter function with the slope that depends on $\epsff$. We find that runs with higher $\epsff$ have the CIMF slope closer to $-2$, which best matches the observations of young star clusters.

\item During a major merger, the CIMF extends to higher masses and has a shallower slope, thus producing more massive clusters than during quiescent epochs.

\item The fraction of clustered star formation $\Gamma$ correlates with $\SigmaSFR$, on a scale of $1\,$kpc. The normalization of this correlation depends strongly on $\epsff$, and the runs with high $\epsff$ best match the observational measurements.

\item The cluster formation timescale depends strongly on $\epsff$ for clusters less massive than $10^5\Msun$. For more massive clusters, the timescale extends to about 3~Myr, the time when the first SNe explode. Runs with $\epsff\leq 0.1$ show a tail of even longer timescales, which is inconsistent with the available measurements of the age spread in young clusters.

\item The derived constraints on $\epsff$ apply to our implementation of the cluster formation algorithm on the spatial scale $3-6\,$pc, and may not correspond directly to the observationally determined efficiency of star formation. We find that for a fixed $\epsff$, the variation in the accretion rate during cluster formation leads to a scatter of the integral efficiency $\epsint$ of at least 0.25~dex. The two runs that best match all observational constraints have median $\epsint \approx 0.16$.
\end{itemize}

\acknowledgements
We thank Eric Bell, Andrey Kravtsov, Diederik Kruijssen, Vadim Semenov, and Anil Seth for helpful discussions. We thank the anonymous referee for valuable suggestions that helped to improve the manuscript.
This work used the Extreme Science and Engineering Discovery Environment (XSEDE) Comet supercomputer at SDSC through allocation AST170007.
We also thank the Michigan Data Science Team (MDST) and XSEDE for providing us computing resources in the high-performance computing center Flux, which was used to perform most of the simulations in this paper. MDST data computation is supported by Advanced Research Computing - Technology Services at the University of Michigan and NVIDIA.
This work was supported in part by NASA through grant NNX12AG44G and by NSF through grant 1412144.

\makeatletter\@chicagotrue\makeatother

\bibliographystyle{apj}
\bibliography{references}

\end{document}